\documentclass[a4paper,12pt]{article}      
\usepackage{jheppub}
\usepackage{slashed}
\usepackage{ulem}
\usepackage[utf8]{inputenc}
\usepackage[T1]{fontenc} 

\begin{document}

\title{Bounding the Dimension-5 Seesaw Portal with Non-Pointing Photon Searches}

\author[a]{L.~Duarte,}
\author[b]{J. Jones-P\'erez,}
\author[b]{C. Manrique-Chavil}

\affiliation[a]{Instituto de F\'{\i}sica, Facultad de Ciencias,
 Universidad de la Rep\'ublica \\ Igu\'a 4225,(11400) 
Montevideo, Uruguay.}
\affiliation[b]{Secci\'on F\'isica, Departamento de Ciencias, Pontificia Universidad Cat\'olica del Per\'u, Apartado 1761, Lima, Peru.}

\emailAdd{lucia@fisica.edu.uy}
\emailAdd{jones.j@pucp.edu.pe}
\emailAdd{cristian.manrique@pucp.edu.pe}

\abstract{
The addition of $d=5$ operators to the Seesaw model leads to the Dimension-5 Seesaw Portal. Here, two new operators provide interactions for the heavy sterile neutrinos. In particular, the Higgs boson can have a large branching ratio into two heavy neutrinos, meaning that these states can be searched for at the LHC. Moreover, the heavy neutrinos can now decay dominantly into light neutrinos and photons. If the heavy neutrinos are long-lived, then searches for delayed, non-pointing photons can constrain the model. In this work, we carry out a detailed recast of an ATLAS search for such displaced photons, triggered by a charged lepton produced in association to the Higgs, placing bounds on the branching ratio for Higgs decay into two heavy neutrinos as low as $2\%$. }

\maketitle
\flushbottom

\section{Introduction}

The Standard Model can be extended with sterile neutrinos $\nu_{Rs}$ in order to account for the light neutrino mass pattern measured in oscillation experiments. The simplest model providing also an explanation of the mass hierarchy between the observed light neutrino states $\nu_{\ell}$ and the other fermion masses, generated only by the Yukawa interactions, is the Seesaw mechanism \cite{Minkowski:1977sc, Mohapatra:1979ia, Yanagida:1980xy, GellMann:1980vs, Schechter:1980gr}. In its simplest versions, it is enough to include two sterile states, which have a Yukawa interaction with SM leptons as well as a Majorana mass term, to obtain a realistic phenomenology. The diagonalization of the mass terms gives the light neutrino states $\nu_\ell$, as well as heavier states $N_h$ that interact with the SM particles via their mixing with the active left-handed states.

The Seesaw is on its own a renormalizable UV complete theory, in the same sense as the SM is. However, when considering right-handed neutrinos at the electroweak scale, one can take it as a low energy effective field theory (EFT), extended with higher dimensional effective operators built from the SM and the right-handed neutrino fields. This theory is currently known as $\nu_R$SMEFT, with a Lagrangian written as:
\begin{equation}
 \mathcal L=\mathcal L_{SM}+\mathcal L_{Seesaw}+ \sum_{d>4} \frac{\alpha_{\mathcal{J}}}{\Lambda^d} \, \mathcal{O}^{d}_{\mathcal{J}}+ h.c.~,
\end{equation}
where the operators $\mathcal{O}_{\mathcal{J}}$ are Lorentz and gauge invariant, and $\Lambda$ is the new physics scale. This EFT, with operators known up to dimension $d=9$~\cite{delAguila:2008ir, Aparici:2009fh, Liao:2016qyd, Bhattacharya:2015vja, Li:2021tsq}, leads to a very rich phenomenology, which has been thoroughly studied in the recent years \cite{Duarte:2015iba,Duarte:2016miz,Duarte:2016caz,Caputo:2017pit,Duarte:2018xst,Yue:2018hci,Duarte:2018kiv,Duarte:2019rzs,Bischer:2019ttk,Alcaide:2019pnf,Butterworth:2019iff,Jones-Perez:2019plk,Chala:2020vqp,Dekens:2020ttz,Barducci:2020ncz,Duarte:2020vgj,Biekotter:2020tbd,DeVries:2020jbs,Barducci:2020icf,Dekens:2021qch,Cirigliano:2021peb,Cottin:2021lzz,Beltran:2021hpq,Zhou:2021ylt,Zhou:2021lnl,Beltran:2022ast,Delgado:2022fea,Barducci:2022gdv,Zapata:2022qwo,Barducci:2022hll,Mitra:2022nri,Fernandez-Martinez:2023phj,Beltran:2023ymm, Zapata:2023wsz,Beltran:2023ksw,Gunther:2023vmz}.
Even though the new interactions are suppressed by the scale $\Lambda$, it is well known that the Seesaw mixing is strongly constrained~\cite{Deppisch:2015qwa,Antusch:2016ejd,Abdullahi:2022jlv}. Thus, it is reasonable to expect that the interactions coming from $\mathcal{O}_{\mathcal{J}}$ can be comparable, if not more important, than those coming from the Seesaw.

In this work we focus on the Dimension-5 Seesaw Portal, that is, the Seesaw model extended with $d=5$ effective operators $\mathcal{O}_{N \phi}$ and $\mathcal{O}_{NB}$. The new interactions provide, in particular, a new pair-production mode from an exotic decay of the Higgs boson $H\to N_h N_{h'}$, and a dipole interaction between the sterile neutrino states and the photon, which are both the subject of our present study. 

The neutrino dipole portal $\mathcal{O}_{NB}$ has caught renovated attention since its proposal as an explanation of the LSND and MiniBooNE anomalies \cite{Kamp:2022bpt, Abdullahi:2023ejc}. Many recent works explore bounds on the neutrino dipole interactions from the intensity, energy and cosmic frontiers, among them  \cite{Magill:2018jla, Schwetz:2020xra, Brdar:2020quo, Brdar:2023tmi,Ovchynnikov:2022rqj}. In particular, in our last work \cite{Delgado:2022fea} we reviewed existing constraints on the neutrino dipole coupling, and re-evaluated the LEP bounds from  $e^{+}e^{-} \to N_h \,\nu$ production. We found that the dipole coupling could not be constrained, unless the mixing was enhanced many orders of magnitude above its naive Seesaw value. In addition, future searches with the capacity of probing the dipole portal have been studied recently at future lepton and hadron colliders \cite{Ovchynnikov:2023wgg, Zhang:2022spf, Zhang:2023nxy}, long-lived particle detectors at the LHC \cite{Liu:2023nxi, Barducci:2022gdv} and from meson decays at the HL-LHC \cite{Barducci:2023hzo}, as well as at neutrino telescopes \cite{Huang:2022pce}.

On the other hand, Higgs decay involving heavy Majorana neutrinos was originally calculated in \cite{Pilaftsis:1991ug}. Attempts to probe it have been carried out for both prompt~\cite{BhupalDev:2012zg,Cely:2012bz} and long-lived heavy neutrinos~\cite{Gago:2015vma}. Moreover, the authors in \cite{Das:2017zjc} used the LHC Higgs data to derive constraints on electroweak-scale sterile Dirac neutrinos. This decay has also received attention in extended models, again for prompt~\cite{Maiezza:2015lza} and long-lived particles~\cite{Deppisch:2018eth, Accomando:2016rpc}.

In the context of the Dimension-5 Seesaw Portal, estimations of the LHC sensitivity reach to test the operator $\mathcal{O}_{N\phi}$ are given in \cite{Caputo:2017pit,Jones-Perez:2019plk}, where the authors consider $pp \to H\to N_h N_h$ for different triggers, with the subsequent displaced decay of the heavy neutrinos into $l^{\pm} q q'$ via the Seesaw mixing. Moreover, in \cite{Barducci:2020icf}, the authors calculate the projected sensitivities of various future Higgs factories to the Seesaw mixings and the branching ratio ${\rm BR}(H\to N_h N_h)$ using similar processes. This sensitivity is also studied at the HL-LHC in~\cite{Mason:2019okp}, with the use of delayed electron signals.

To the best of our knowledge, the combined presence of both operators has not been studied using LHC data in order to place bounds on the model. In this work, we extend our results in~\cite{Delgado:2022fea} and focus on a scenario where the heavy neutrinos are pair-produced via the Higgs portal $\mathcal{O}_{N \phi}$, and can decay to a photon and a light neutrino via the dipole portal $\mathcal{O}_{NB}$. In the regime where the heavy neutrinos $N_h$ are long-lived, this leads to final states with photons which are displaced (delayed and non-pointing), a remarkable signal which has been recently searched for at the LHC~\cite{ATLAS:2022vhr} (see also \cite{ATLAS:2013etx,ATLAS:2014kbb,CMS:2012bbi,CMS:2019zxa,ATLAS:2023meo}). By carefully recasting this search in terms of the Dimension-5 Seesaw Portal interactions, bounds are placed, for the first time, on the coefficient of the $\mathcal{O}_{N \phi}$ operator, based on an existing LHC result. These bounds are applicable provided that the dipole interaction strength allows for the $N_h$ to be long-lived but still decay inside the detector, disintegrating primarily into final states with photons. 

The paper is organised as follows. In Section~\ref{sec:overview}, we review the model and present the associated processes that will be critical for the search. In Section~\ref{sec:bounds} we provide a detailed description of our recast of the search for delayed photons, report our statistical analysis, and show our final results. We finally conclude in Section~\ref{sec:conclusions}.

\section{Overview of the Dimension-5 Seesaw Portal}
\label{sec:overview}

The standard Seesaw model, with two sterile neutrinos $\nu_R$, has the following Lagrangian:
\begin{equation}
 \mathcal L=\mathcal L_{SM}+i\bar\nu_{Rs}\,\slashed\partial\,\nu_{Rs}-\left(\bar L_a (Y_\nu)_{a s}\,\tilde\phi\, \nu_{Rs}
 +\frac{1}{2}\bar\nu_{Rs} (M_N)_{ss'}\nu^c_{Rs'}+h.c.\right)~,
\end{equation}
where $a=e,\,\mu,\,\tau$ and $s,s'=s_1,s_2$. The $Y_\nu$ couplings are connected to Dirac masses, which appear on the neutrino mass matrix alongside the symmetric Majorana mass $M_N$. As is well known, after diagonalization two non-zero masses for the SM neutrinos can be obtained\footnote{Note that a third non-zero mass for the light neutrinos can be generated, without changing our results, by adding an additional sterile neutrino, which can be later decoupled~\cite{Jones-Perez:2019plk}.}. Light states are thus denoted $\nu_\ell$ ($\ell=1,\,2,\,3$), with masses $m_\ell$, while heavy states are denoted $N_h$ ($h=4,\,5$), with masses $M_h$. Heavy neutrinos can be probed through their active neutrino component, as parametrized in the mixing matrix $U$, which allows them to interact via the $W$ and $Z$ bosons. The latest bounds on these mixings can be found in~\cite{Blennow:2023mqx, Abdullahi:2022jlv}.

In the following, we extend the Seesaw with the following $d=5$ operators\footnote{We assume that the Weinberg operator~\cite{Weinberg:1979sa} gives a negligible contribution to the light neutrino mass matrix.}:
 \begin{equation}
 \label{eq:LagEff}
  \mathcal L_5=\frac{(\alpha_{N\phi})_{ss'}}{\Lambda}(\phi^\dagger\phi)\,\bar\nu_{Rs}\,\nu^c_{Rs'}
  +\frac{(\alpha_{NB})_{ss'}}{\Lambda}\bar\nu_{Rs}\,\sigma^{\mu\nu}\nu^c_{Rs'}\,B_{\mu\nu}+h.c.~,
 \end{equation}
where $\sigma^{\mu\nu}=\tfrac{i}{2}[\gamma^\mu,\,\gamma^\nu]$. These are referred to as Anisimov-Graesser $\mathcal{O}_{N \phi}$~\cite{Anisimov:2006hv,Graesser:2007yj,Graesser:2007pc} and dipole $\mathcal{O}_{NB}$~\cite{Aparici:2009fh} operators, respectively. The former involves a symmetric coefficient $\alpha_{N\phi}$, while the latter has an antisymmetric $\alpha_{NB}$. It is important to point out that, although denoted as $\alpha_{N\phi}/\Lambda$ and $\alpha_{NB}/\Lambda$, the new physics scale $\Lambda$ does not necessarily have to be the same.

The inclusion of these operators modifies the neutrino mass matrix, as well as their couplings. Following~\cite{Delgado:2022fea}, we will neglect all effects on the masses, and concentrate exclusively on the new interactions. In the following, we list all interaction terms relevant for our work. First, the Anisimov-Graesser operator allows the coupling of the Higgs to two heavy neutrinos:
\begin{equation}
 \label{eq:L_higgsnn}
\mathcal L_h =
\frac{v}{\Lambda}H\, \bar N_h\left[(\alpha^{\prime*}_{N\phi})_{hh'}P_R+(\alpha^{\prime}_{N\phi})_{hh'}P_L\right]\,N_{h'}~,
\end{equation}
with $(\alpha'_{N\phi})_{hh'} = U_{sh}\,(\alpha_{N\phi})_{ss'}\,U_{s'h'}$, having $U_{sh}\sim I$ as the ``sterile-heavy'' mixing sector of $U$. In principle, the Higgs can also couple to two heavy neutrinos through the $Y_\nu$ coupling in the standard Seesaw. This, however, is very strongly suppressed, as in addition to the smallness of $Y_\nu$, the coupling also requires an ``active-heavy'' mixing, $U_{ah}\sim\sqrt{m_\ell/M_h}$.

The dipole operator allows the heavy neutrinos to interact with the photon and $Z$ bosons:
\begin{eqnarray}
\label{eq:L_Zhnn}
\mathcal L_{ZNN} &=& -\frac{s_W}{\Lambda}(\partial_\mu Z_\nu-\partial_\nu Z_\mu)\,\bar N_4\,\sigma^{\mu\nu}\left[(\alpha'_{NB})_{4 5}P_L-(\alpha^{\prime\,*}_{NB})_{4 5}P_R\right]N_5 +h.c.~, \\
\label{eq:L_gammahnn}
\mathcal L_{\gamma NN} &=& \frac{c_W}{\Lambda}(\partial_\mu A_\nu-\partial_\nu A_\mu)\,\bar N_4\,\sigma^{\mu\nu}\left[(\alpha'_{NB})_{4 5}P_L-(\alpha^{\prime\,*}_{NB})_{4 5}P_R\right]N_5 + h.c. ~.
\end{eqnarray}
As in the previous case, we define $(\alpha'_{NB})_{45} = U_{s4}\,(\alpha_{NB})_{ss'}\,U_{s'5}$. Given that the coupling is antisymmetric, the heavy neutrinos involved must necessarily be different. It is worth noting that, similarly to the Higgs, a coupling with the $Z$ is also allowed by the standard Seesaw, but is also heavily suppressed.

Finally, we report the coupling of one heavy neutrino, a light neutrino (or charged lepton) and a gauge boson:
\begin{eqnarray}
 \mathcal L_W&=&\frac{g}{\sqrt2} W_\mu^-\bar\ell_a\gamma^\mu\, U_{a h}\,P_L\, N_h+h.c. ~,\\
 \mathcal L_Z &=&\frac{g}{4c_W} Z_\mu \bar \nu_\ell\gamma^\mu\, \left(C_{\ell h}\,P_L-C_{\ell h}^*\,P_R\right)\, N_h \nonumber \\
 &&-\frac{s_W}{\Lambda}(\partial_\mu Z_\nu-\partial_\nu Z_\mu)\,\bar \nu_\ell\,\sigma^{\mu\nu}\left[(\alpha'_{NB})_{\ell h}P_L-(\alpha^{\prime\,*}_{NB})_{\ell h}P_R\right]N_h +h.c. ~, \label{eq:L_Znn}\\
 \mathcal L_\gamma &=&\frac{c_W}{\Lambda}(\partial_\mu A_\nu-\partial_\nu A_\mu)\,\bar \nu_\ell\,\sigma^{\mu\nu}\left[(\alpha'_{NB})_{\ell h}P_L-(\alpha^{\prime\,*}_{NB})_{\ell h}P_R\right]N_h  + h.c. ~,\label{eq:L_gammann}
\end{eqnarray}
with $C_{\ell h}= U^*_{a \ell}\,U_{a h}$ and $(\alpha'_{NB})_{\ell h} = U_{s\ell}\,(\alpha_{NB})_{ss'}\,U_{s'h}$. These are always suppressed, either by $U_{ah}$ or by the sterile-light mixing, $U_{s\ell}\sim\sqrt{m_\ell/M_h}$.

\subsection{Production and Decay Channels at the LHC}

We consider heavy neutrino pair production from Higgs decays. In our model, this decay is dominated by the couplings appearing in Eq.~(\ref{eq:L_higgsnn}). Taking them as real, the corresponding partial width is~\cite{Graesser:2007yj,Aparici:2009fh}:
\begin{equation}
\label{eq:HiggsDecay}
\Gamma(H\to N_h\,N_{h'}) =
S_{hh'}\frac{v^2}{2\pi}
\frac{\sqrt{\lambda(m_H^2,\,M_h^2,\,M_{h'}^2)}}{m_H}
\left|\frac{(\alpha'_{N\phi})_{hh'}}{\Lambda}\right|^2
\left(1-\frac{(M_h+M_{h'})^2}{m_H^2}\right)~.
\end{equation}
Here, $m_H$ is the Higgs mass, $\lambda(x,\,y,\,z)= x^2+y^2+z^2-2xy-2xz-2yz$, and $S_{hh'}$ provides a factor $1/2$ if the two outgoing heavy neutrinos are the same.

The decay of the Higgs into two $N_{4}$, two $N_5$ or an $N_4\,N_5$ pair depends on the structure of the couplings $\alpha'_{N\phi}$. As presented in~\cite{Caputo:2017pit,Delgado:2022fea}, in order to explain the size difference between the coefficients of the Weinberg operator and both $\mathcal{O}_{N \phi}$ and $\mathcal{O}_{NB}$ operators, one could argue the existence of a slightly broken lepton number symmetry. From this, one would expect the diagonal elements of $\alpha_{N\phi}$ to be very suppressed, meaning that $H\to N_4\,N_5$ would be favoured. In what follows, we will consider this decay exclusively.

Since we are taking $M_4<m_H/2$, the lightest heavy neutrino $N_4$ will either decay into a light neutrino and a photon, or into a variety of final states described by three-body decays. The $\nu_\ell\,\gamma$ decay is mediated by $\alpha_{NB}$, and was first calculated in~\cite{Aparici:2009fh}:
\begin{equation}
 \label{eq:decayintogamma}
 \Gamma(N_4\to\,\nu\,\gamma)=\frac{2}{\pi}\,c_W^2 M_4^3 \sum_\ell\left|\frac{(\alpha'_{NB})_{\ell 4}}{\Lambda}\right|^2~,
\end{equation}
where $c_W$ is the cosine of the weak mixing angle. Since this decay mode requires the heavily suppressed sterile-light mixing, the width is small, allowing for $N_4$ to be long-lived.

The calculation of three-body decays are somewhat more complicated, as the mixing with interaction states leads to diagrams with virtual $W^\pm$ and $Z$ bosons (see~\cite{Atre:2009rg,Kovalenko:2009td,Helo:2010cw,Bondarenko:2018ptm,Coloma:2020lgy} for the corresponding widths in the standard Seesaw). As was shown in~\cite{Delgado:2022fea}, it is very important to include three-body decays in order to estimate correctly the heavy neutrino decay length. In addition, it is also possible to introduce new contributions involving a virtual $Z$ or photon, however, these have a very small impact on the branching ratios and decay lengths within the region of the parameter space we are interested in. The full formulae can be found in Appendix~B of~\cite{Delgado:2022fea}.

If the Higgs decays into two heavy neutrinos of different mass, the heaviest is expected to decay via either $N_5\to N_4\,\gamma$ or $N_5\to N_4\,Z$, as their couplings are not suppressed by mixing. For simplicity, in the following we will restrict ourselves to $\Delta M\equiv M_5-M_4< M_Z$, so the latter is forbidden. In this case, the width of the heavy neutrino is given by~\cite{Aparici:2009fh}:
\begin{equation}
\label{eq:heavydecayintogamma}
\Gamma(N_5\to\,N_4\,\gamma)=\frac{2}{\pi}\,c_W^2 \frac{(M_5^2-M_4^2)^3}{M^3_5}\left|\frac{(\alpha'_{NB})_{4 5}}{\Lambda}\right|^2~,
\end{equation}
Notice that, compared to Eq.~(\ref{eq:decayintogamma}), this width depends on $(\alpha'_{NB})_{h h'}$ instead of $(\alpha'_{NB})_{\ell h}$. As mentioned before, this implies that the decay will proceed without the need of sterile-light mixing, so it will not be suppressed. Thus, apart from rendering the $N_5$ as short-lived, three body decays are irrelevant when calculating the $N_5$ width.

An exception to this expectation might arise if the heavy neutrinos are pseudo-Dirac particles, with practically degenerate masses. As for the dimension-5 operator hierarchy, this possibility can be expected in the presence of an approximate lepton number symmetry~\cite{Branco:1988ex,Shaposhnikov:2006nn,Kersten:2007vk,Gavela:2009cd,Hernandez:2018cgc}, which is a feature of several Seesaw realizations~\cite{Wyler:1982dd,Mohapatra:1986bd,Malinsky:2005bi,Kang:2006sn}. Then, decays of $N_5$ exclusively into SM particles, such as the one shown in Eq.~(\ref{eq:decayintogamma}), can dominate if:
\begin{equation}
\label{eq:how_degen}
\left(1-\frac{M^2_4}{M^2_5}\right)\ll |U_{s\ell}|^2~.
\end{equation}
For instance, as shown in~\cite{Antusch:2017ebe,Fernandez-Martinez:2022gsu}, in certain models the heavy mass splitting could be as low as the light neutrino solar mass splitting, implying that the term on the left-hand side of Eq.~(\ref{eq:how_degen}) would be of order $10^{-14}$, for GeV masses. Thus, if the mixing on the right-hand side was larger than this, it would be necessary to calculate both two and three body decays for $N_5$, as is done for $N_4$. In this work we will assume that the mass splitting is always large enough such that this never happens. In particular, we assume that the mixing is never above the naive Seesaw expectation, which maximises the lifetime of $N_4$.

\section{Bounds from Non-Pointing Photon Searches}
\label{sec:bounds}

The ATLAS search for non-pointing photons~\cite{ATLAS:2022vhr} used $\mathcal L=139$~fb$^{-1}$ of data collected from proton-proton collisions at a center-of-mass energy of $\sqrt s=13$~TeV. It is assumed that the photons come from the decays of a pair of LLPs, generated in turn by the decay of the Higgs boson. The measurement is triggered by a prompt electron or muon, with $p_T>27$~GeV, coming from associated production with the Higgs.

As in the search, our simulation considered the triggering leptons coming from $p\,p\to W^\pm H$, $p\,p\to Z\,H$ and $p\,p\to t\,\bar t\, H$ processes. Events were generated with \texttt{MadGraph5\_aMC@NLO 2.9.7}~\cite{Alwall:2014hca}, which uses \texttt{LHAPDF6}~\cite{Buckley:2014ana}. The model was the same of~\cite{Delgado:2022fea}, based on~\texttt{FeynRules 2.3.43}~\cite{Christensen:2008py,Alloul:2013bka}. The heavy neutrino decay, parton showering and hadronization was carried out by~\texttt{PYTHIA 8.244}~\cite{Sjostrand:2006za}, giving a \texttt{HepMC} file as output~\cite{Dobbs:2001ck,Buckley:2019xhk}. The cross-sections, calculated at leading order, were multiplied by appropriate K-factors, following~\cite{LHCHiggsCrossSectionWorkingGroup:2016ypw}.

\subsection{Arrival Time and Non-Pointing Parameter}
\label{sec:arr.nonpoint}

The analysis in~\cite{ATLAS:2022vhr} relies on two kinematical variables. The first is the time delay $t_\gamma$, that is, the difference between the arrival time and that expected from a prompt photon. The second one is the non-pointing parameter $|\Delta z_\gamma|$, defined as the distance between the interaction point and the extrapolated trajectory of the reconstructed photon, measured along the beamline. In our work, both $t_\gamma$ and $|\Delta z_\gamma|$ are initially calculated for all photons from truth-level information, extracted from the \texttt{HepMC} output of \texttt{PYTHIA}, which assumes collisions happening at $t=0$ and $x=y=z=0$. These are smeared at a later stage, in order to take into account the experimental resolution, as we explain in Section~\ref{sec:Ev.Selection}. Our procedure was first outlined in~\cite{Delgado:2022fea}, which we reproduce here for the convenience of the reader.

In order to determine $t_\gamma$, we calculated the absolute time $t'$ for each photon to enter the ECal, which depends on the heavy neutrino momentum $\vec p_N$, its decay position $\vec r_N$, and the direction of the photon momentum $\vec p_\gamma$ after the decay. The region where each photon entered the ECal would be mapped to a specific pseudorapidity $\eta$, measured with respect to the center of the detector. Similarly, it is possible to calculate the corresponding time $t_0$ related to prompt photons, also as a function of $\eta$. With these two, the time delay is simply $t_\gamma=t'-t_0$.

On the other hand, $|\Delta z_\gamma|$ was obtained by calculating the point on the beam axis closest to the extrapolated trajectory of the photon (see Appendix C of~\cite{FrancoThesis}). After some simplifications, one finds\footnote{The ATLAS search does not use the $\phi$ information of the measured energy deposits, meaning that these are projected on the $z-R$ plane. By taking two points, corresponding to deposits in distinct ECal layers, the photon trajectory on this plane is reconstructed. This adds an extra term on Eq.~(\ref{eq:DeltaZ}), which depends on $R_1$ and $R_2$ coordinates associated to each layer (see Chapter~7 in~\cite{Nikiforou:2014cka}, and Figure~5 in~\cite{ATLAS:2009zsq}). We have checked that this additional term does not add significant modifications, and is thus dropped.}:
\begin{eqnarray}
|\Delta z_\gamma|=
r_{N_z}-\frac{p_{\gamma_z}}{{p^2_\gamma}_T}(r_{N_x}\,p_{\gamma_x}+r_{N_y}\,p_{\gamma_y})~,
\label{eq:DeltaZ}
\end{eqnarray}
where we have assumed that the primary vertex lies at the center of the detector. The variables $r_{N_i}$ and $p_{\gamma_i}$ denote the components of $\vec r_N$ and $\vec p_\gamma$, and ${p_\gamma}_T$ is the photon transverse momentum.

Once this information was obtained, it was stored within the same \texttt{HepMC} file, and passed on to \texttt{Delphes} in order to carry out the detector simulation, including photon timing and pointing smearing.

\subsection{Event Reconstruction in Delphes}

The event reconstruction is simulated by \texttt{Delphes 3.5.0}~\cite{deFavereau:2013fsa}, which depends on \texttt{FastJet 3.4.0}~\cite{Cacciari:2011ma}, with the exception of lepton and photon ID efficiencies, as well as overlap removal, implemented at a later stage in our analysis. We modified the code, such that it would read the $t_\gamma$ and $|\Delta z_\gamma|$ information stored within the \texttt{HepMC} file, and later include it in its output.

The detector simulation includes most modules from the ATLAS card packaged within \texttt{Delphes}, with appropriate modifications. To begin with, the \texttt{ParticlePropagator} module was modified in order to give more details of the ATLAS dimensions~\cite{ATLAS:2008xda}, so, apart from the magnetic field coverage of $1.15\,$m, and a half-length of $3.512\,$m, we also specified an ECal inner radius of $1.5\,$m (\texttt{RadiusMax}$=1.5$).

For photons, we left all calorimetry modules untouched, returning \texttt{EFlow} photons with reconstructed ${p_\gamma}_T$. Following~\cite{ATLAS:2022vhr}, we modified the photon track isolation, such that it required the scalar sum of $p_T$ of all tracks within $\Delta R=0.2$ of the photon, with $p_T>1\,$GeV, to be less than $5\%$ of ${p_\gamma}_T$. In addition, we added a photon calorimeter isolation module, which required the sum of all energy deposits on the ECal, within $\Delta R=0.2$, to be less than $6.5\%$ of ${p_\gamma}_T$. The photon ID efficiency was set to $100\%$, with the real efficiency to be applied outside of \texttt{Delphes} (see below).

For electron reconstruction we proceeded almost identically as for photons. The electron track isolation restricted the sum of $p_T$ of additional tracks around the electron to be less than $15\%$ of the reconstructed electron transverse momentum, ${p_e}_T$. The calorimeter isolation would then require other energy deposits to be less than $20\%$ of ${p_e}_T$. Finally, the electron ID efficiency was also to be implemented outside \texttt{Delphes}.

For muons, we eliminated the isolation module and set a $100\%$ ID efficiency, with the objective of applying both outside \texttt{Delphes}. For jets, we set the $R$ parameter to $0.4$, and a minimum $p_T$ of $25\,$GeV. Finally, we deactivated the \texttt{UniqueObjectFinder} module, as the overlap removal was to be implemented after the ID efficiencies had been applied, based on the specific requirements in~\cite{ATLAS:2022vhr}. As commented earlier, the \texttt{Delphes} output routines were also modified, such that the $t_\gamma$ and $|\Delta z_\gamma|$ variables were written for each reconstructed photon.

\subsection{Event Selection}\label{sec:Ev.Selection}

As noted above, the event reconstructed in our \texttt{Delphes} implementation is incomplete. In the following, we describe how we finalise the detector simulation, and furthermore give details on the event selection and analysis performed in~\cite{ATLAS:2022vhr}.

The photons selected for the analysis must satisfy basic cuts, such as having $p_T>10\,$GeV and $\eta<2.37$ (with the region between the ECal barrel and endcaps excluded), as well as being generated within the Inner Detector, before the ECal. In order to take into account the experimental resolution, both $t_\gamma$ and $|\Delta z_\gamma|$ variables were each Gaussian smeared\footnote{It must be noted that the smearing of $t_\gamma$ takes into account the timing spread of $pp$ collision times.}. For $|\Delta z_\gamma|$, the resolution for the smear was taken from an interpolation of the data shown in Figure 1 of~\cite{ATLAS:2014kbb}. For $t_\gamma$, we interpolated the data in Figure 2 of~\cite{ATLAS:2022vhr}. The latter Figure is presented as a function of $E_{\rm cell}$, defined as the middle-layer ECal cell receiving the maximum energy deposit of the shower. For definiteness, following the claim in~\cite{ATLAS:2022vhr} of $E_{\rm cell}$ being around $20$--$50\%$ of the total energy deposited by the shower, we set $E_{\rm cell}=0.35\,{E_\gamma}$, where $E_\gamma$ is the total photon energy, as reported by \texttt{Delphes}.

Given the smeared value for $|\Delta z_\gamma|$, we applied the photon ID efficiency shown in Figure 3 of~\cite{ATLAS:2022vhr}. Then, following the procedure of the search, at least one photon was required to be in the ECal barrel. Events satisfying this constraint were divided into single or multi-photon channels. In the case of having more than one photon in the barrel, the one with the largest ${p_\gamma}_T$ was used for the analysis. For this photon, a further cut $E_{\rm cell}>10\,$GeV was imposed.

Regarding charged leptons, only those with $p_T>10\,$GeV would be considered in the analysis. Electrons would be restricted to $\eta<2.47$, again excluding the region between the ECal barrel and endcap, while muons would be required to have $\eta<2.7$. For the electron ID efficiency, we followed the data within both panels in Figure 17 of~\cite{ATLAS:2019qmc}, for medium electrons. Furthermore, for muon isolation and ID efficiency, we followed the total identification efficiency curve in Figure 21 of~\cite{ATLAS:2020auj}.

Turning to jets, apart from the $p_T>25\,$GeV cut outlined above, we also asked for a rapidity $|y|<4.4$. Having defined photons, electrons, muons and jets, we proceeded with the overlap removal in~\cite{ATLAS:2022vhr}. Thus, all electrons with $\Delta R\leq0.4$ from a photon were removed from the event. Then, all jets within $\Delta R\leq0.4$ from a photon, or $\Delta R\leq0.2$ from an electron, were removed. Next, in order to match the requirements appearing in the measurement of isolated electron efficiencies, all electrons within $\Delta R\leq0.4$ from the surviving jets were removed. Finally, all muons within $\Delta R\leq0.4$ from photons or jets were also removed.

With these unique, reconstructed objects in hand, the search required one charged lepton to match the triggering lepton. In other words, one of the surviving electrons or muons must have $p_T>27\,$GeV. If the triggering lepton is an electron, then the invariant mass of the electron - photon pair, $m_{e\gamma}$, must also satisfy $|m_{e\gamma}-m_Z|>15\,$GeV. With this, the event is selected for analysis.

In order to be assigned to the signal region, the missing transverse energy (MET) must be larger than 50~GeV. Events in the signal region are then classified into five categories, depending on the value of $|\Delta z_\gamma|$. Within each category, events are binned following the value of $t_\gamma$. The binning depends on whether the event sample is in the single or multi-photon channel.

We have validated our recast by implementing the signal model of~\cite{ATLAS:2022vhr}, and generating the timing distributions for all $|\Delta z_\gamma|$ categories, for both channels\footnote{We thank S.~N.~Santpur for providing the \texttt{param\_card} files for \texttt{MadGraph}.}. The resulting events in each category were consistent in order of magnitude and in behaviour with respect to $t_\gamma$, compared to Figure 7 of~\cite{ATLAS:2022vhr}.

\subsection{Statistical Analysis}

We now turn to the statistical analysis used to constrain our model. We remind the reader that the long-lived heavy neutrinos are produced via $H\to N_4\,N_5$ decay, which in turn disintegrate into final states with photons. The constraints on the total number of signal events can then be translated as bounds on the Higgs branching ratio into $N_4\,N_5$, which can then be expressed in terms of $(\alpha'_{N\phi})_{45}/\Lambda$ using  Eq.~\eqref{eq:HiggsDecay}. Our objective is to constrain this parameter as a function of the heavy neutrino mass and lifetime, the latter depending on $\alpha_{NB}/\Lambda$.

To this end, we use the model independent results presented by ATLAS in~\cite{ATLAS:2022vhr}, which focuses on the last timing $t_\gamma$ bin of the category with largest non-pointing $|\Delta z_\gamma|$, for single and multi-photon channels. The number of measured events in such bins, as well as the backgrounds, are given in Table II of~\cite{ATLAS:2022vhr}. Specifically, the single (multi-) photon channel uses a $t_\gamma$ bin between 1.5 and 12 ns (1 and 12 ns), observing 4 (0) events, while expecting $3.8\pm1.6$ ($0.28\pm0.04$). Both channels focus on $|\Delta z_\gamma|>300\,$mm. The Table also presents a ``combination'' of both channels, with 4 events observed and $4.1\pm1.7$ expected. Notice, however, that the latter actually corresponds to a merging of the information in single and multi-photon channels into a single bin, rather than a statistical combination of two separate bins.

Our 95\% exclusion limits are calculated using the CLs method. This method generally overcovers the confidence interval, with the objective of avoiding setting bounds on signal rates which the experiment is insensitive to. This occurs because it takes into account possible background underfluctuations~\cite{Read:2002hq}, and thus gives softer bounds than the CL method. Our implementation follows the PDG review on Statistics \cite{ParticleDataGroup:2022pth} and Apprendix B in \cite{Magill:2018jla}. We consider a counting experiment with a Poisson likelihood function, and calculate the upper number of signal events $s^{up}$ consistent at $(1 - \alpha)=95\%$ confidence level with the observation of $n$ events and a background prediction of $b$ events. This is done solving for $s^{up}$ in the following equation
\begin{equation}
    \alpha'= \frac{\alpha}{(1-\alpha_b)}= 0.05~,
\end{equation}
where
\begin{equation}
    \alpha= e^{-s^{up}} \left( \frac{\sum_{m=0}^{m=n} (s^{up}+ b)^{m}/ m!}{\sum_{m=0}^{m=n} b^{m}/ m!}  \right) \qquad \text{and}  \qquad \alpha_b= \int_{n}^{\infty} \frac{b^{\nu} }{\nu!} e^{-b} d\nu~.
\end{equation}
For the $n$ and $b$ values specified above, we obtain $s^{up}= 6.8$ ($3.8$) for the single (multi\nobreakdash) photon channels, and $s^{up}= 6.8$ for the combination. This allows us to exclude the parameter space region in which the interpolated predicted number of signal events $s$ exceeds this value, where:
\begin{eqnarray}
s=\mathcal L\,\sum_{X=Z,\,W,\,t\bar t}K_X\,\sigma_{XH}\,{\rm BR}(H\to N_4\,N_5)\frac{N^{\rm cuts}_X}{N^{\rm gen}_X}~.
\end{eqnarray}
In the equation above, $\mathcal L$ is the integrated luminosity, $\sigma_{XH}$ is the leading-order cross-section for associated Higgs production, as provided by \texttt{MadGraph}, $K_X$ is the corresponding K-factor~\cite{LHCHiggsCrossSectionWorkingGroup:2016ypw}, $N_X^{\rm gen}$ are the number of events generated for each production process, and $N_X^{\rm cuts}$ are the corresponding number of events surviving the cuts. The $N_4$ and $N_5$ branching ratio information is included in $N_X^{\rm cuts}$, for each value of $M_h$ and $(\alpha_{NB})/\Lambda$ to be evaluated. Thus, a bound in $s$ can be interpreted as a bound in ${\rm BR}(H\to N_4\,N_5)$. In the following, we will use the combined channel analysis to constrain this branching ratio.

In order to verify this implementation of the CLs method, we used the same events generated for our validation, based on the signal model of the search, to reproduce the bounds shown in Figure 11 of~\cite{ATLAS:2022vhr}. Our curves were consistent again in order of magnitude and general behaviour.

\subsection{Results}

We are now in condition of presenting the results of our analysis. We explore two benchmark scenarios, with mass differences $\Delta M= M_5-M_4 =1~\text{GeV}$ and $\Delta M= \, 15~\text{GeV}$. In both, the Higgs decays into $N_5$ and $N_4$ pairs, with the $N_5$ then disintegrating into a prompt photon and an $N_4$. The final pair of $N_4$ will each usually decay into a displaced photon and a light neutrino.

\begin{figure}[tbp]
\includegraphics[width=0.49\textwidth]{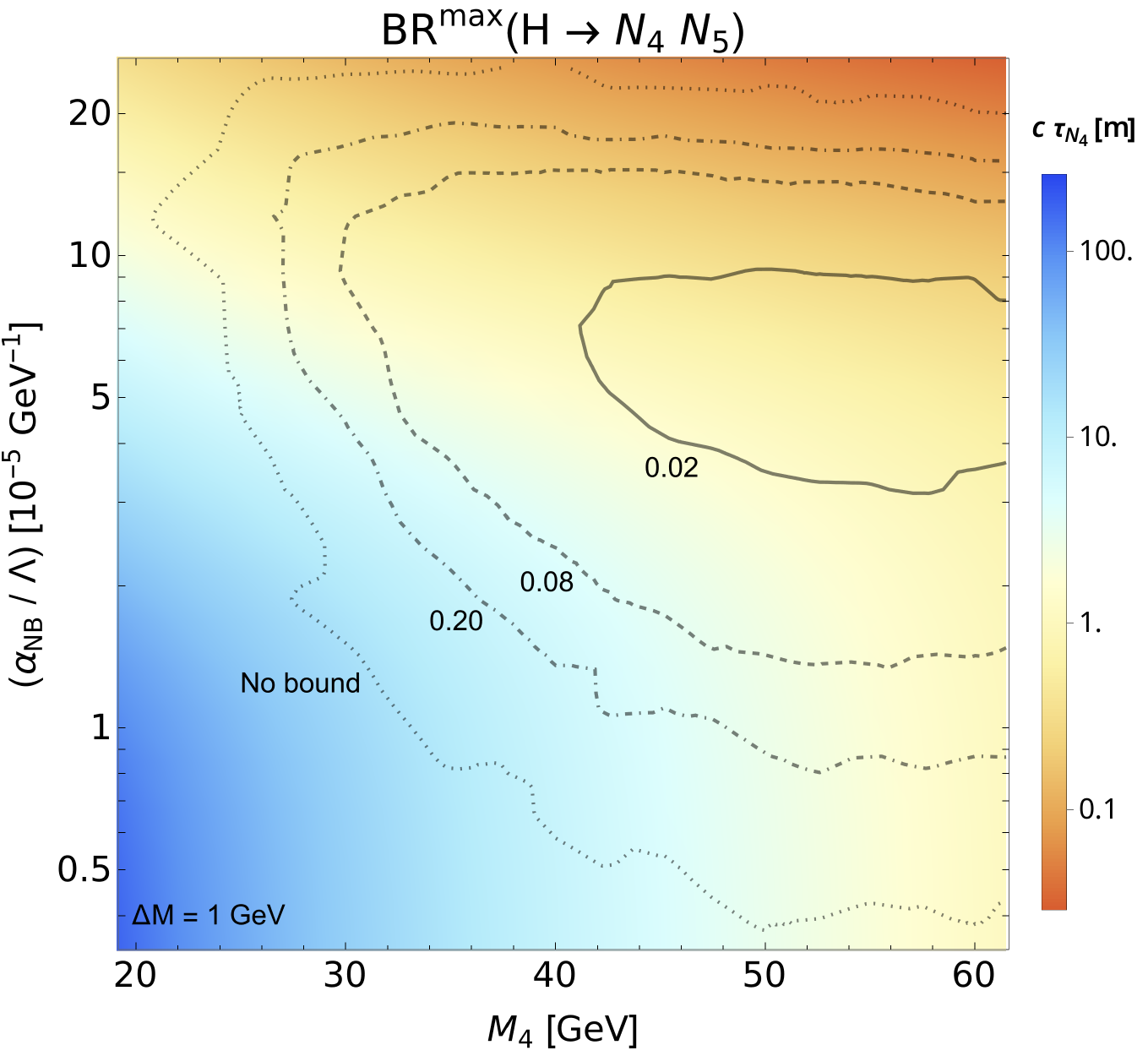} \hfill
\includegraphics[width=0.435\textwidth]{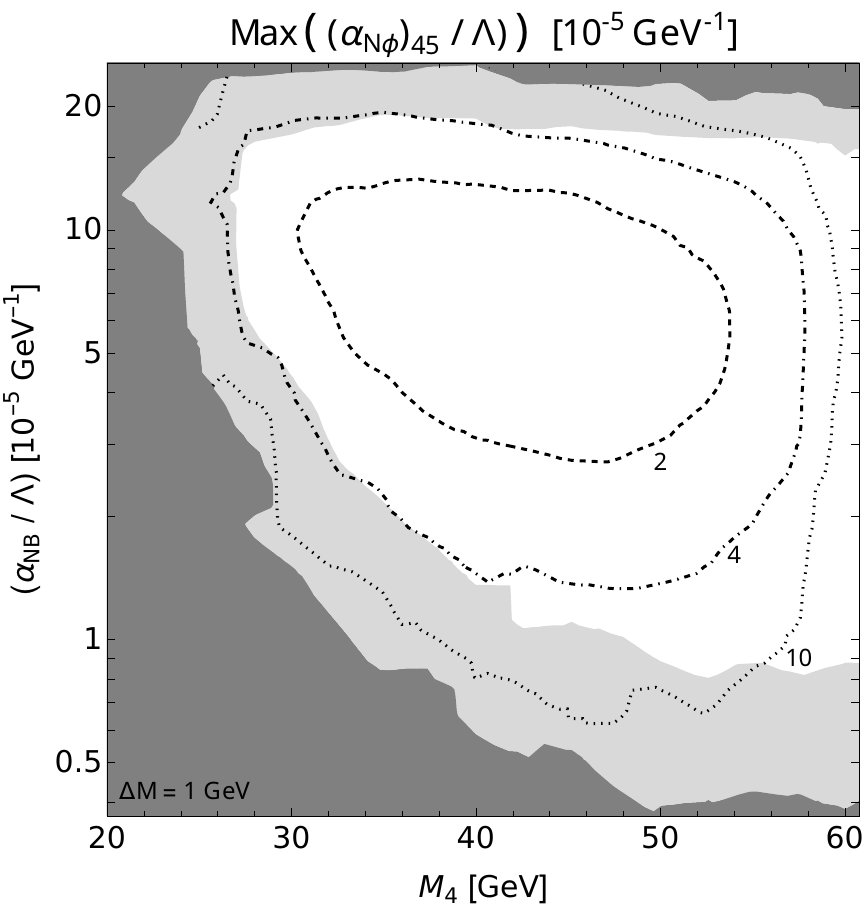}
\caption{Left: Bounds on ${\rm BR}(H\to N_4\,N_5)$, for $\Delta M=1\,$GeV. The coloured background gives the magnitude of the decay length of $N_4$. Right: Bounds on $(\alpha'_{N\phi})_{45}/\Lambda$, for $\Delta M=1\,$GeV. Dark grey regions have no bound, while the light gray regions imply branching ratios larger than $20\%$.} 
\label{fig:Bounds1}
\end{figure}
In our first benchmark scenario, we consider $\Delta M=1 \,$GeV. In this case, the prompt photon will carry very little $p_T$, so is unlikely to pass the selection cuts. We thus consider this benchmark as representative of all cases with very small $\Delta M$, as well as those with $H\to N_4\,N_4$ decays. Results are plotted in Figure~\ref{fig:Bounds1}. The left panel of the Figure shows our bounds on ${\rm BR}(H\to N_4\,N_5)$, as contours on the $M_4 - \alpha_{NB}/\Lambda$ plane\footnote{In this Section we denote $\alpha_{NB}\equiv(\alpha_{NB})_{s_1 s_2}$.}. We find that the search is sensitive to the model for $M_4$ between 20 and 60~GeV, and $\alpha_{NB}/\Lambda$ between $0.4\times10^{-5} \,\text{GeV}^{-1}$ and $2.2\times10^{-4}\,$GeV$^{-1}$, approximately. In particular, we find a region where the branching ratio is constrained to be under $20\%$, which is below the bounds from ATLAS and CMS searches for Higgs decays to exotic undetected states~\cite{CMS:2018uag,ATLAS:2019nkf}.

One can easily understand the sensitivity curves. First, from the point of view of $\alpha_{NB}/\Lambda$, values under $\sim10^{-5}\,$GeV$^{-1}$ can have heavy neutrinos with a too long lifetime, such that most of them escape the detector without leaving a signal. In contrast, for values larger than $\sim10^{-4}\,$GeV$^{-1}$, the $N_4$ decays much more promptly, which is reflected in our events being categorized in the smaller $|\Delta z_\gamma|$ and $t_\gamma$ bins, and thus not counted in the statistical analysis.

Moreover, for a fixed $\alpha_{NB}$, a small $M_4$ corresponds to a larger lifetime, increasing the possibility for the $N_4$ to escape the detector without leaving a signal. A large $M_4$, in contrast, also implies that the heavy neutrinos are produced with smaller velocity. Having a slow-moving parent leads to the photons being more delayed, which increases the number of events in the large $t_\gamma$ bin used in the analysis. Furthermore, the lower the momenta, the less collimated each neutrino - photon pair are, which favours having a large $|\Delta z_\gamma|$. However, in some regions of the parameter space, the mass cannot be arbitrarily large, as the $N_4\to\nu\,\gamma$ branching ratio can become smaller than unity, decreasing the overall final number of events.

On the left panel of the Figure we also show the decay length of $N_4$, with colours ranging from red ($c\tau\leq0.1\,$m) to blue ($c\tau\geq100\,$m). The (red) region with the smallest decay length corresponds to large $M_4$ and $\alpha_{NB}/\Lambda$, while the opposite side of the panel corresponds to the largest one (blue). The boundary between red and blue regions (in white) corresponds to a decay length around 1~m. Interestingly, for large $\alpha_{NB}/\Lambda$ this boundary crosses the plot diagonally, until it reaches values $\alpha_{NB}/\Lambda\lesssim3\times10^{-5}\,$GeV$^{-1}$. At this point, the boundary twists downwards, and proceeds vertically. This means that, within this region, the dipole operator is no longer the dominant contribution for the decay length, having a large component from the standard Seesaw partial width. Moreover, in this region the $N_h\to\nu\,\gamma$ branching ratio is strongly suppressed~\cite{Delgado:2022fea}.

As can be seen in Eq.~(\ref{eq:HiggsDecay}), for given $M_4,\,M_5$, the bounds on the Higgs branching ratios can be translated to constraints on the Anisimov-Graesser coefficient\footnote{For the values of mixing we are using, one can take $(\alpha'_{N\phi})_{45}=(\alpha_{N\phi})_{s_1 s_2}$, given that $U_{sh}\sim I$.}, $(\alpha'_{N\phi})_{45}/\Lambda$. These are shown on the right panel of Figure~\ref{fig:Bounds1}. Here, the dark gray region corresponds to no bound on ${\rm BR}(H\to N_4\,N_5)$, while for the light gray region the limit is above $20\%$, meaning that the searches in~\cite{CMS:2018uag,ATLAS:2019nkf} would give stronger bounds on the coefficient. Within the non-shaded region, $(\alpha'_{N\phi})_{45}/\Lambda$ is limited to values of order $\sim10^{-5}\,$GeV$^{-1}$. Notice that the area that constrains this coefficient most strongly is somewhat displaced to lighter masses with respect to the corresponding region for the branching ratio. The reason for this is that for large values of $M_4~\text{and}~M_5$ the Higgs branching ratio is naturally suppressed due to the smaller phase space, allowing for larger $\alpha'_{N\phi}/\Lambda$.

\begin{figure}[tbp]
\includegraphics[width=0.49\textwidth]{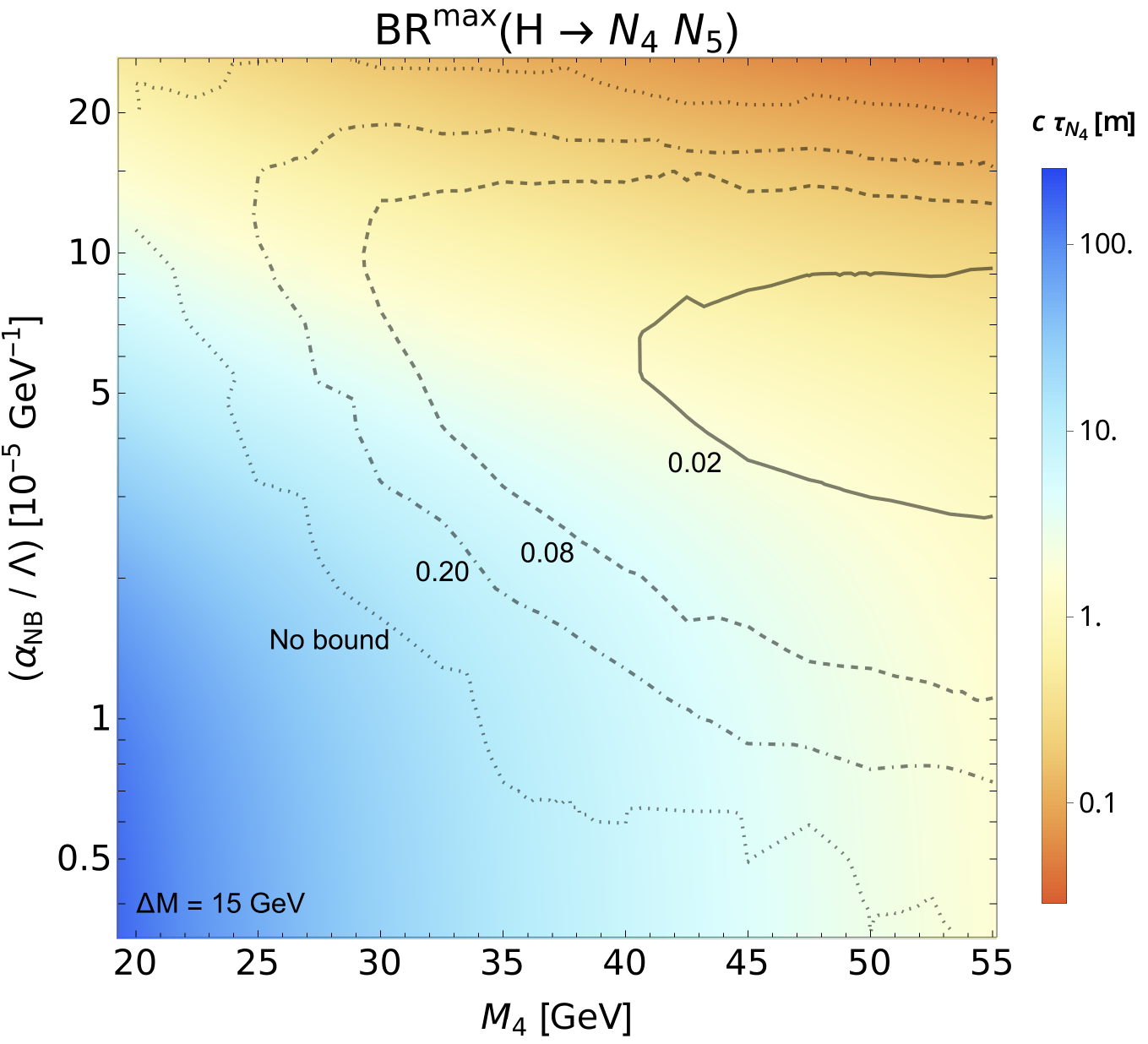} \hfill
\includegraphics[width=0.44\textwidth]{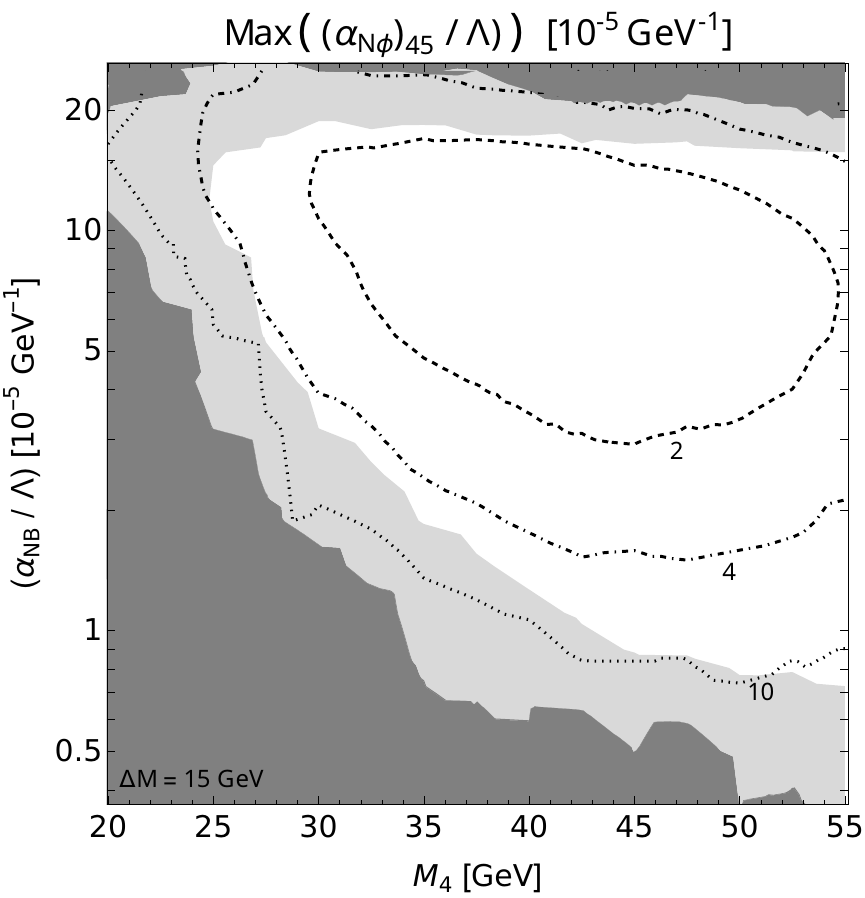}
\caption{Same as Figure~\ref{fig:Bounds1}, but for $\Delta M=15\,$GeV.} 
\label{fig:Bounds15}
\end{figure}
Similarly, Figure~\ref{fig:Bounds15} shows the same bounds, but for $\Delta M=15\,$GeV. This scenario differs from the previous one principally in the fact that the prompt photon from $N_5\to N_4\,\gamma$ decay will have a larger $p_T$, having thus better chances of passing the event selection than the one in the $\Delta M=1\,$GeV scenario. Notice this does not mean that this prompt photon will be used in the analysis, but rather that each event will be more likely to be assigned to the multi-photon channel.

On the left panel, we find that having a larger heavy neutrino mass difference tends to somewhat improve the sensitivity of the search with respect to the Higgs branching ratio, with the exclusion regions reaching smaller values of both $M_4$ and $\alpha_{NB/\Lambda}$. The reason for this is that, for fixed $M_4$, having a larger $M_5$ implies that the heavy neutrinos will have less momentum and, as argued earlier, this is correlated to larger $t_\gamma$ and $|\Delta z_\gamma|$ values. 

On the right panel of Figure~\ref{fig:Bounds15}, we again show the constraints on $(\alpha'_{N\phi})_{45}/\Lambda$. The sensitivity is very similar to that shown in Figure~\ref{fig:Bounds1}, with the regions slightly displaced towards larger values of $\alpha_{NB}/\Lambda$. This displacement implies that for larger $\alpha_{NB}/\Lambda$ the $\Delta M=15\,$GeV scenario is more sensitive than that for $\Delta M=1\,$GeV. However, for intermediate $\alpha_{NB}/\Lambda$, it turns out that the constraints on the branching ratio are slightly stronger for $\Delta M=1\,$GeV

\begin{figure}[tbp]
\centering
\includegraphics[width=0.49\textwidth]{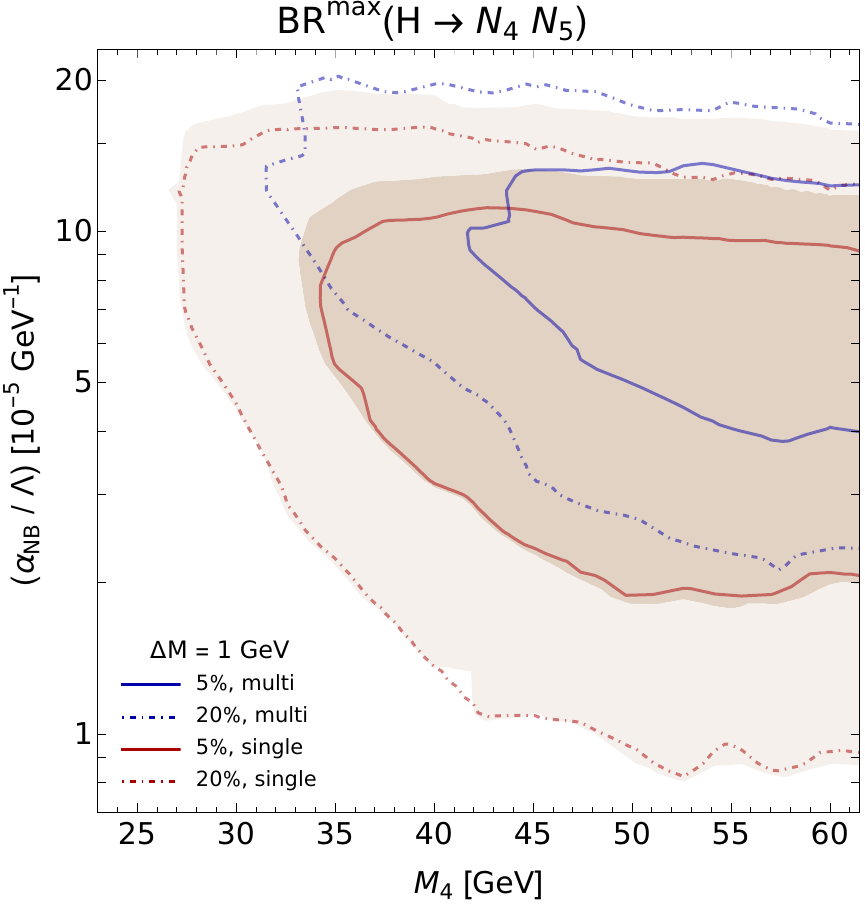} \hfill
\includegraphics[width=0.49\textwidth]{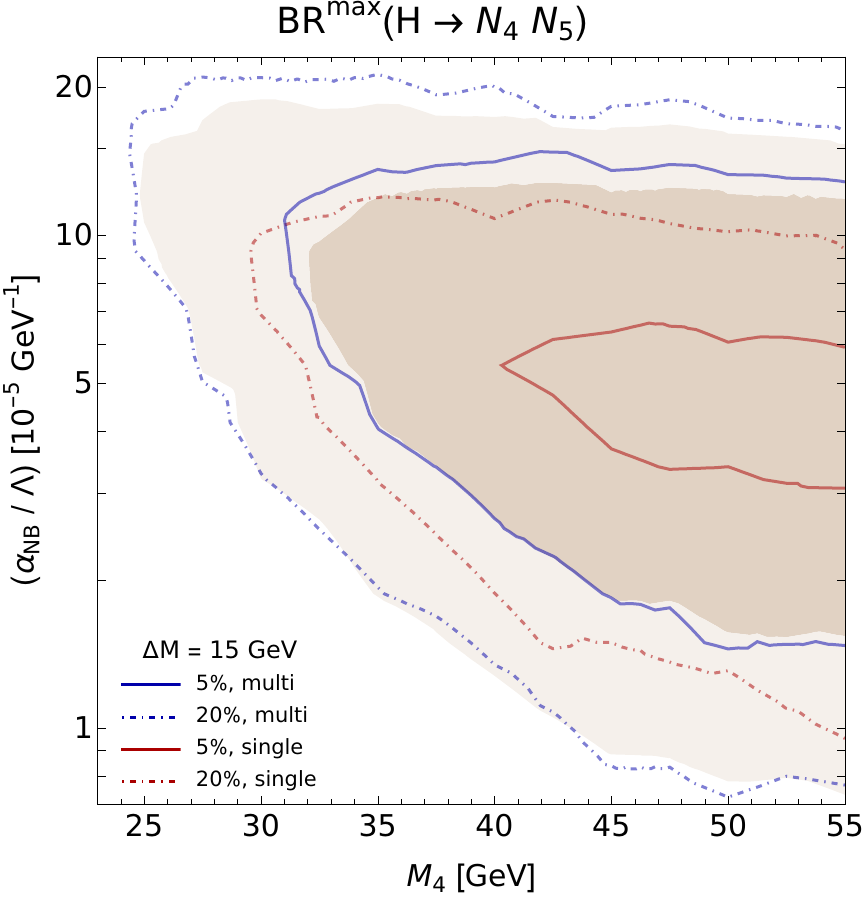}
\caption{Comparison of single and multiphoton channels, shown in red and blue, respectively. Branching ratio limits of $5\%$ ($20\%$) are shown in solid (dashed-dotted) lines. Corresponding bounds using combined channels are shown in dark (light) brown shading.} 
\label{fig:BoundsComparison}
\end{figure}
Perhaps more interestingly, in Figure~\ref{fig:BoundsComparison} we compare the constraints on ${\rm BR}(H\to N_4\,N_5)$ coming from the combined channel analysis, with those obtained separately from the single and multi-photon channels. The limits from the combined channel are shown as brown regions, while those from the single (multi-) photon channel are shown as red (blue) lines.

The scenario with $\Delta M=1\,$GeV is shown on the left. As expected, the region probed by the combined analysis comes from the union of regions probed by the individual channels. In particular, the multi-photon channel is most sensitive to regions with large $\alpha_{NB}/\Lambda$, while the single photon channel determines the sensitivity at low values. This is reasonable, the smaller $\alpha_{NB}/\Lambda$ is, the larger the chances that a heavy neutrino will escape the detector before decaying, removing one displaced photon from the analysis. Another reason for losing displaced photons at small $\alpha_{NB}/\Lambda$ is the suppressed $N_4\to\,\nu\,\gamma$ branching ratio.

It is interesting to note that in some regions the multi-photon channel is more sensitive than the combination. We attribute this, on the one hand, to the multi-photon channel having essentially no background, leading to a large sensitivity, and, on the other hand, as mentioned earlier, to the combined analysis not including the per-bin information but just adding the events in both channels instead.

Finally, we turn to the right panel, which shows the scenario with $\Delta M=15\,$GeV. As commented before, the prompt photon from $N_5\to N_4\,\gamma$ is more likely to pass the selection. Thus, most events have one additional photon in comparison to $\Delta M=1\,$GeV, transferring displaced single photon events to the multi-photon channel. As commented before, since this channel has a much smaller background, the sensitivity is greater. Correspondingly, we have far fewer events on the single photon channel, leading to much milder constraints on the parameter space. Thus, it makes sense to have the combined analysis following the exclusion set by the multi-photon channel.

Before we conclude, it is important to comment on the interpretation of the results in the context of $\nu_R$SMEFT. As is well known, the effective operators can be generated from both perturbative and non-perturbative extensions\footnote{Also referred respectively as decoupled and strong-interacting regimes.} of the Seesaw Lagrangian~\cite{Aparici:2009fh}. In particular, in perturbative extensions the dipole operator $\mathcal{O}_{NB}$ in Eq.~(\ref{eq:LagEff}) is generated from loop diagrams, which would lead to an extra factor $(4\pi)^{2}$ in the denominator. This leads to a rescaling of $(\alpha_{NB})/\Lambda$ by a factor $\sim158$, meaning that, in a perturbative extension, our bounds would be applicable for $\Lambda$ around 100~GeV. Notice that since $(\alpha_{NB})/\Lambda$ only participates via the decay of the heavy neutrino, the energy scales involving this operator are of order $M_4$, meaning that the $\nu_R$SMEFT can still be applicable.

It is important to note that this does not necessarily mean that new physics beyond the $\nu_R$SMEFT should be expected at the 100~GeV scale. As a simple example of UV-completion, we consider the model in~\cite{Aparici:2009oua}, which adds a scalar $\omega$ and a vector-like fermion $E$, transforming both only under $U(1)_Y$. In this model, for $m_\omega\ll m_E$, the dipole operator is in fact generated with a $1/m_E$ factor, with the implication that $m_E$ should be around 100 GeV. However, if $m_\omega\gg m_E$, the overall factor comes out as $m_E/m_\omega^2$, allowing for much larger masses while keeping $\Lambda$ at low values.

On the other hand, if the effective operators are generated in non-perturbative extensions, then these considerations need not to be taken into account. Thus, the bounds can be taken as shown without the need of rescaling. However, as commented in~\cite{Barducci:2022gdv}, in this scenario an explanation for the difference between $M_{N_h}$ and $\Lambda$ scales is lacking.

\section{Conclusions}
\label{sec:conclusions}

After more than 15 years in operation, the LHC has provided a huge amount of precious information. Searches such as the one studied in this work, probing the existence of LLPs, are a proof of the power and flexibility of this machine. It is thus essential to fully exploit all available data, in order to place bounds on new physics models.

To this end, we performed a detailed recast of the ATLAS search for displaced photons coming from LLP decays, triggered by a charged lepton associated to Higgs production~\cite{ATLAS:2022vhr}. This required the calculation of both non-pointing and time delay variables, $|\Delta z_\gamma|$ and $t_\gamma$. Apart from this, we refined the detector simulation beyond the standard settings in \texttt{Delphes}, implementing the isolation cuts, the resolutions, the efficiencies, and the overlap removal, reported by ATLAS for the search. We also reproduced the model-independent statistical analysis performed by the search.

With this in hand, we explored the sensitivity of the search to the Dimension-5 Seesaw Portal. Apart from the addition of sterile neutrinos, which generate light neutrino masses, this model includes two new $d=5$ effective operators featuring the sterile neutrinos themselves. The Anisimov-Graesser operator $\mathcal{O}_{N \phi}$, described by the parameter $\alpha_{N\phi}/\Lambda$, is an important contribution towards Higgs decay into the two heavy neutrinos. The dipole operator $\mathcal{O}_{N B}$, described by $\alpha_{NB}/\Lambda$, allows the heavy neutrinos to decay into photons and lighter neutrinos. Thus, in the region of parameter space where the heavy neutrinos have long lifetimes, the model can provide the final state the search is sensitive to.

As a result, we find that the search can constrain the branching ratio for Higgs decays into two heavy neutrinos, as long as the latter have masses between 20 and 60~GeV, and $\alpha_{NB}/\Lambda$ lies between $10^{-4}$ and $10^{-6}\,$GeV$^{-1}$. The bounds on the branching ratio can be as low as $2\%$. This in turn can be translated into limits on $\alpha_{N\phi}/\Lambda$, which can be constrained to values as small as $2\times10^{-5}\,$GeV$^{-1}$.

In our work we also compare the sensitivity of the various channels used in the search. We find that, in general, the one photon channel is most sensitive to regions where the LLP lifetime is large, while the multi-photon channel is more appropriate for regions where the lifetime is shorter. However, this statement is modified in the presence of prompt photons generated along the LLPs, giving a much more important role to the multi-photon channel.

\acknowledgments
We would like to thank D.~Mahon and S.~N.~Santpur for discussions. We acknowledge the financial support of ProCIENCIA, CONCYTEC (Peru's National Council for Science and Technology), Contract 123-2020-FONDECYT. J.J.P.\ acknowledges funding by the {\it Direcci\'on de Gesti\'on de la Investigaci\'on} at PUCP, through grant DGI-2021-C-0020.  L.D. is funded by PEDECIBA-F\'isica.

\bibliographystyle{utphys}
\bibliography{main}

\providecommand{\href}[2]{#2}\begingroup\raggedright\begin{thebibliography}{100}

\bibitem{Minkowski:1977sc}
P.~Minkowski, ``{$\mu \to e\gamma$ at a Rate of One Out of $10^{9}$ Muon
  Decays?},''
\href{http://dx.doi.org/10.1016/0370-2693(77)90435-X}{{\em Phys. Lett.}
  {\bfseries B67} (1977) 421--428}.

\bibitem{Mohapatra:1979ia}
R.~N. Mohapatra and G.~Senjanovic, ``{Neutrino Mass and Spontaneous Parity
  Violation},''
\href{http://dx.doi.org/10.1103/PhysRevLett.44.912}{{\em Phys. Rev. Lett.}
  {\bfseries 44} (1980) 912}.

\bibitem{Yanagida:1980xy}
T.~Yanagida, ``{Horizontal Symmetry and Masses of Neutrinos},''
\href{http://dx.doi.org/10.1143/PTP.64.1103}{{\em Prog.Theor.Phys.} {\bfseries
  64} (1980) 1103}.

\bibitem{GellMann:1980vs}
M.~Gell-Mann, P.~Ramond, and R.~Slansky, ``{Complex Spinors and Unified
  Theories},'' {\em Conf. Proc.} {\bfseries C790927} (1979) 315--321,
\href{http://arxiv.org/abs/1306.4669}{{\ttfamily arXiv:1306.4669 [hep-th]}}.

\bibitem{Schechter:1980gr}
J.~Schechter and J.~W.~F. Valle, ``{Neutrino Masses in SU(2) x U(1)
  Theories},''
\href{http://dx.doi.org/10.1103/PhysRevD.22.2227}{{\em Phys. Rev.} {\bfseries
  D22} (1980) 2227}.

\bibitem{delAguila:2008ir}
F.~del Aguila, S.~Bar-Shalom, A.~Soni, and J.~Wudka, ``{Heavy Majorana
  Neutrinos in the Effective Lagrangian Description: Application to Hadron
  Colliders},'' \href{http://dx.doi.org/10.1016/j.physletb.2008.11.031}{{\em
  Phys.Lett.} {\bfseries B670} (2009) 399--402},
\href{http://arxiv.org/abs/0806.0876}{{\ttfamily arXiv:0806.0876 [hep-ph]}}.

\bibitem{Aparici:2009fh}
A.~Aparici, K.~Kim, A.~Santamaria, and J.~Wudka, ``{Right-handed neutrino
  magnetic moments},'' \href{http://dx.doi.org/10.1103/PhysRevD.80.013010}{{\em
  Phys. Rev. D} {\bfseries 80} (2009) 013010},
  \href{http://arxiv.org/abs/0904.3244}{{\ttfamily arXiv:0904.3244 [hep-ph]}}.

\bibitem{Liao:2016qyd}
Y.~Liao and X.-D. Ma, ``{Operators up to Dimension Seven in Standard Model
  Effective Field Theory Extended with Sterile Neutrinos},''
  \href{http://dx.doi.org/10.1103/PhysRevD.96.015012}{{\em Phys. Rev.}
  {\bfseries D96} no.~1, (2017) 015012},
\href{http://arxiv.org/abs/1612.04527}{{\ttfamily arXiv:1612.04527 [hep-ph]}}.

\bibitem{Bhattacharya:2015vja}
S.~Bhattacharya and J.~Wudka, ``{Dimension-seven operators in the standard
  model with right handed neutrinos},''
  \href{http://dx.doi.org/10.1103/PhysRevD.94.055022,
  10.1103/PhysRevD.95.039904}{{\em Phys. Rev.} {\bfseries D94} no.~5, (2016)
  055022}, \href{http://arxiv.org/abs/1505.05264}{{\ttfamily arXiv:1505.05264
  [hep-ph]}}.
[Erratum: Phys. Rev.D95,no.3,039904(2017)].

\bibitem{Li:2021tsq}
H.-L. Li, Z.~Ren, M.-L. Xiao, J.-H. Yu, and Y.-H. Zheng, ``{Operator bases in
  effective field theories with sterile neutrinos: d \ensuremath{\leq} 9},''
  \href{http://dx.doi.org/10.1007/JHEP11(2021)003}{{\em JHEP} {\bfseries 11}
  (2021) 003}, \href{http://arxiv.org/abs/2105.09329}{{\ttfamily
  arXiv:2105.09329 [hep-ph]}}.

\bibitem{Duarte:2015iba}
L.~Duarte, J.~Peressutti, and O.~A. Sampayo, ``{Majorana neutrino decay in an
  Effective Approach},''
  \href{http://dx.doi.org/10.1103/PhysRevD.92.093002}{{\em Phys. Rev.}
  {\bfseries D92} no.~9, (2015) 093002},
\href{http://arxiv.org/abs/1508.01588}{{\ttfamily arXiv:1508.01588 [hep-ph]}}.

\bibitem{Duarte:2016miz}
L.~Duarte, I.~Romero, J.~Peressutti, and O.~A. Sampayo, ``{Effective Majorana
  neutrino decay},''
  \href{http://dx.doi.org/10.1140/epjc/s10052-016-4301-8}{{\em Eur. Phys. J. C}
  {\bfseries 76} no.~8, (2016) 453},
  \href{http://arxiv.org/abs/1603.08052}{{\ttfamily arXiv:1603.08052
  [hep-ph]}}.

\bibitem{Duarte:2016caz}
L.~Duarte, J.~Peressutti, and O.~A. Sampayo, ``{Not-that-heavy Majorana
  neutrino signals at the LHC},''
  \href{http://dx.doi.org/10.1088/1361-6471/aa99f5}{{\em J. Phys. G} {\bfseries
  45} no.~2, (2018) 025001}, \href{http://arxiv.org/abs/1610.03894}{{\ttfamily
  arXiv:1610.03894 [hep-ph]}}.

\bibitem{Caputo:2017pit}
A.~Caputo, P.~Hernandez, J.~Lopez-Pavon, and J.~Salvado, ``{The seesaw portal
  in testable models of neutrino masses},''
  \href{http://dx.doi.org/10.1007/JHEP06(2017)112}{{\em JHEP} {\bfseries 06}
  (2017) 112}, \href{http://arxiv.org/abs/1704.08721}{{\ttfamily
  arXiv:1704.08721 [hep-ph]}}.

\bibitem{Duarte:2018xst}
L.~Duarte, G.~Zapata, and O.~A. Sampayo, ``{Angular and polarization trails
  from effective interactions of Majorana neutrinos at the LHeC},''
  \href{http://dx.doi.org/10.1140/epjc/s10052-018-5833-x}{{\em Eur. Phys. J.}
  {\bfseries C78} no.~5, (2018) 352},
\href{http://arxiv.org/abs/1802.07620}{{\ttfamily arXiv:1802.07620 [hep-ph]}}.

\bibitem{Yue:2018hci}
C.-X. Yue and J.-P. Chu, ``{Sterile neutrino and leptonic decays of the
  pseudoscalar mesons},''
  \href{http://dx.doi.org/10.1103/PhysRevD.98.055012}{{\em Phys. Rev. D}
  {\bfseries 98} no.~5, (2018) 055012},
  \href{http://arxiv.org/abs/1808.09139}{{\ttfamily arXiv:1808.09139
  [hep-ph]}}.

\bibitem{Duarte:2018kiv}
L.~Duarte, G.~Zapata, and O.~A. Sampayo, ``{Final taus and initial state
  polarization signatures from effective interactions of Majorana neutrinos at
  future $e^{+}e^{-}$ colliders},''
  \href{http://dx.doi.org/10.1140/epjc/s10052-019-6734-3}{{\em Eur. Phys. J.}
  {\bfseries C79} no.~3, (2019) 240},
\href{http://arxiv.org/abs/1812.01154}{{\ttfamily arXiv:1812.01154 [hep-ph]}}.

\bibitem{Duarte:2019rzs}
L.~Duarte, J.~Peressutti, I.~Romero, and O.~A. Sampayo, ``{Majorana neutrinos
  with effective interactions in B decays},''
  \href{http://dx.doi.org/10.1140/epjc/s10052-019-7104-x}{{\em Eur. Phys. J.}
  {\bfseries C79} no.~7, (2019) 593},
\href{http://arxiv.org/abs/1904.07175}{{\ttfamily arXiv:1904.07175 [hep-ph]}}.

\bibitem{Bischer:2019ttk}
I.~Bischer and W.~Rodejohann, ``{General neutrino interactions from an
  effective field theory perspective},''
  \href{http://dx.doi.org/10.1016/j.nuclphysb.2019.114746}{{\em Nucl. Phys. B}
  {\bfseries 947} (2019) 114746},
  \href{http://arxiv.org/abs/1905.08699}{{\ttfamily arXiv:1905.08699
  [hep-ph]}}.

\bibitem{Alcaide:2019pnf}
J.~Alcaide, S.~Banerjee, M.~Chala, and A.~Titov, ``{Probes of the Standard
  Model effective field theory extended with a right-handed neutrino},''
  \href{http://dx.doi.org/10.1007/JHEP08(2019)031}{{\em JHEP} {\bfseries 08}
  (2019) 031}, \href{http://arxiv.org/abs/1905.11375}{{\ttfamily
  arXiv:1905.11375 [hep-ph]}}.

\bibitem{Butterworth:2019iff}
J.~M. Butterworth, M.~Chala, C.~Englert, M.~Spannowsky, and A.~Titov, ``{Higgs
  phenomenology as a probe of sterile neutrinos},''
  \href{http://dx.doi.org/10.1103/PhysRevD.100.115019}{{\em Phys. Rev. D}
  {\bfseries 100} no.~11, (2019) 115019},
  \href{http://arxiv.org/abs/1909.04665}{{\ttfamily arXiv:1909.04665
  [hep-ph]}}.

\bibitem{Jones-Perez:2019plk}
J.~Jones-P\'erez, J.~Masias, and J.~Ruiz-\'Alvarez, ``{Search for Long-Lived
  Heavy Neutrinos at the LHC with a VBF Trigger},''
  \href{http://dx.doi.org/10.1140/epjc/s10052-020-8188-z}{{\em Eur. Phys. J. C}
  {\bfseries 80} no.~7, (2020) 642},
  \href{http://arxiv.org/abs/1912.08206}{{\ttfamily arXiv:1912.08206
  [hep-ph]}}.

\bibitem{Chala:2020vqp}
M.~Chala and A.~Titov, ``{One-loop matching in the SMEFT extended with a
  sterile neutrino},'' \href{http://dx.doi.org/10.1007/JHEP05(2020)139}{{\em
  JHEP} {\bfseries 05} (2020) 139},
  \href{http://arxiv.org/abs/2001.07732}{{\ttfamily arXiv:2001.07732
  [hep-ph]}}.

\bibitem{Dekens:2020ttz}
W.~Dekens, J.~de~Vries, K.~Fuyuto, E.~Mereghetti, and G.~Zhou, ``{Sterile
  neutrinos and neutrinoless double beta decay in effective field theory},''
  \href{http://dx.doi.org/10.1007/JHEP06(2020)097}{{\em JHEP} {\bfseries 06}
  (2020) 097}, \href{http://arxiv.org/abs/2002.07182}{{\ttfamily
  arXiv:2002.07182 [hep-ph]}}.

\bibitem{Barducci:2020ncz}
D.~Barducci, E.~Bertuzzo, A.~Caputo, and P.~Hernandez, ``{Minimal flavor
  violation in the see-saw portal},''
  \href{http://dx.doi.org/10.1007/JHEP06(2020)185}{{\em JHEP} {\bfseries 06}
  (2020) 185}, \href{http://arxiv.org/abs/2003.08391}{{\ttfamily
  arXiv:2003.08391 [hep-ph]}}.

\bibitem{Duarte:2020vgj}
L.~Duarte, G.~Zapata, and O.~Sampayo, ``{Angular and polarization observables
  for Majorana-mediated B decays with effective interactions},''
  \href{http://dx.doi.org/10.1140/epjc/s10052-020-08471-0}{{\em Eur. Phys. J.
  C} {\bfseries 80} no.~9, (2020) 896},
  \href{http://arxiv.org/abs/2006.11216}{{\ttfamily arXiv:2006.11216
  [hep-ph]}}.

\bibitem{Biekotter:2020tbd}
A.~Biek\"otter, M.~Chala, and M.~Spannowsky, ``{The effective field theory of
  low scale see-saw at colliders},''
  \href{http://dx.doi.org/10.1140/s10052-020-8339-2}{{\em Eur. Phys. J. C}
  {\bfseries 80} no.~8, (2020) 743},
  \href{http://arxiv.org/abs/2007.00673}{{\ttfamily arXiv:2007.00673
  [hep-ph]}}.

\bibitem{DeVries:2020jbs}
J.~De~Vries, H.~K. Dreiner, J.~Y. G\"unther, Z.~S. Wang, and G.~Zhou,
  ``{Long-lived Sterile Neutrinos at the LHC in Effective Field Theory},''
  \href{http://dx.doi.org/10.1007/JHEP03(2021)148}{{\em JHEP} {\bfseries 03}
  (2021) 148}, \href{http://arxiv.org/abs/2010.07305}{{\ttfamily
  arXiv:2010.07305 [hep-ph]}}.

\bibitem{Barducci:2020icf}
D.~Barducci, E.~Bertuzzo, A.~Caputo, P.~Hernandez, and B.~Mele, ``{The see-saw
  portal at future Higgs Factories},''
  \href{http://dx.doi.org/10.1007/JHEP03(2021)117}{{\em JHEP} {\bfseries 03}
  (2021) 117}, \href{http://arxiv.org/abs/2011.04725}{{\ttfamily
  arXiv:2011.04725 [hep-ph]}}.

\bibitem{Dekens:2021qch}
W.~Dekens, J.~de~Vries, and T.~Tong, ``{Sterile neutrinos with non-standard
  interactions in \ensuremath{\beta}- and
  0\ensuremath{\nu}\ensuremath{\beta}\ensuremath{\beta}-decay experiments},''
  \href{http://dx.doi.org/10.1007/JHEP08(2021)128}{{\em JHEP} {\bfseries 08}
  (2021) 128}, \href{http://arxiv.org/abs/2104.00140}{{\ttfamily
  arXiv:2104.00140 [hep-ph]}}.

\bibitem{Cirigliano:2021peb}
V.~Cirigliano, W.~Dekens, J.~de~Vries, K.~Fuyuto, E.~Mereghetti, and R.~Ruiz,
  ``{Leptonic anomalous magnetic moments in \ensuremath{\nu} SMEFT},''
  \href{http://dx.doi.org/10.1007/JHEP08(2021)103}{{\em JHEP} {\bfseries 08}
  (2021) 103}, \href{http://arxiv.org/abs/2105.11462}{{\ttfamily
  arXiv:2105.11462 [hep-ph]}}.

\bibitem{Cottin:2021lzz}
G.~Cottin, J.~C. Helo, M.~Hirsch, A.~Titov, and Z.~S. Wang, ``{Heavy neutral
  leptons in effective field theory and the high-luminosity LHC},''
  \href{http://dx.doi.org/10.1007/JHEP09(2021)039}{{\em JHEP} {\bfseries 09}
  (2021) 039}, \href{http://arxiv.org/abs/2105.13851}{{\ttfamily
  arXiv:2105.13851 [hep-ph]}}.

\bibitem{Beltran:2021hpq}
R.~Beltr\'an, G.~Cottin, J.~C. Helo, M.~Hirsch, A.~Titov, and Z.~S. Wang,
  ``{Long-lived heavy neutral leptons at the LHC: four-fermion single-N$_{R}$
  operators},'' \href{http://dx.doi.org/10.1007/JHEP01(2022)044}{{\em JHEP}
  {\bfseries 01} (2022) 044}, \href{http://arxiv.org/abs/2110.15096}{{\ttfamily
  arXiv:2110.15096 [hep-ph]}}.

\bibitem{Zhou:2021ylt}
G.~Zhou, J.~Y. G\"unther, Z.~S. Wang, J.~de~Vries, and H.~K. Dreiner,
  ``{Long-lived sterile neutrinos at Belle II in effective field theory},''
  \href{http://dx.doi.org/10.1007/JHEP04(2022)057}{{\em JHEP} {\bfseries 04}
  (2022) 057}, \href{http://arxiv.org/abs/2111.04403}{{\ttfamily
  arXiv:2111.04403 [hep-ph]}}.

\bibitem{Zhou:2021lnl}
G.~Zhou, ``{Light sterile neutrinos and lepton-number-violating kaon decays in
  effective field theory},''
  \href{http://dx.doi.org/10.1007/JHEP06(2022)127}{{\em JHEP} {\bfseries 06}
  (2022) 127}, \href{http://arxiv.org/abs/2112.00767}{{\ttfamily
  arXiv:2112.00767 [hep-ph]}}.

\bibitem{Beltran:2022ast}
R.~Beltr\'an, G.~Cottin, J.~C. Helo, M.~Hirsch, A.~Titov, and Z.~S. Wang,
  ``{Long-lived heavy neutral leptons from mesons in effective field theory},''
  \href{http://dx.doi.org/10.1007/JHEP01(2023)015}{{\em JHEP} {\bfseries 01}
  (2023) 015}, \href{http://arxiv.org/abs/2210.02461}{{\ttfamily
  arXiv:2210.02461 [hep-ph]}}.

\bibitem{Delgado:2022fea}
F.~Delgado, L.~Duarte, J.~Jones-Perez, C.~Manrique-Chavil, and S.~Pe\~na,
  ``{Assessment of the dimension-5 seesaw portal and impact of exotic Higgs
  decays on non-pointing photon searches},''
  \href{http://dx.doi.org/10.1007/JHEP09(2022)079}{{\em JHEP} {\bfseries 09}
  (2022) 079}, \href{http://arxiv.org/abs/2205.13550}{{\ttfamily
  arXiv:2205.13550 [hep-ph]}}.

\bibitem{Barducci:2022gdv}
D.~Barducci, E.~Bertuzzo, M.~Taoso, and C.~Toni, ``{Probing right-handed
  neutrinos dipole operators},''
  \href{http://dx.doi.org/10.1007/JHEP03(2023)239}{{\em JHEP} {\bfseries 03}
  (2023) 239}, \href{http://arxiv.org/abs/2209.13469}{{\ttfamily
  arXiv:2209.13469 [hep-ph]}}.

\bibitem{Zapata:2022qwo}
G.~Zapata, T.~Urruzola, O.~A. Sampayo, and L.~Duarte, ``{Lepton collider probes
  for Majorana neutrino effective interactions},''
  \href{http://dx.doi.org/10.1140/epjc/s10052-022-10448-0}{{\em Eur. Phys. J.
  C} {\bfseries 82} no.~6, (2022) 544},
  \href{http://arxiv.org/abs/2201.02480}{{\ttfamily arXiv:2201.02480
  [hep-ph]}}.

\bibitem{Barducci:2022hll}
D.~Barducci and E.~Bertuzzo, ``{The see-saw portal at future Higgs factories:
  the role of dimension six operators},''
  \href{http://dx.doi.org/10.1007/JHEP06(2022)077}{{\em JHEP} {\bfseries 06}
  (2022) 077}, \href{http://arxiv.org/abs/2201.11754}{{\ttfamily
  arXiv:2201.11754 [hep-ph]}}.

\bibitem{Mitra:2022nri}
M.~Mitra, S.~Mandal, R.~Padhan, A.~Sarkar, and M.~Spannowsky, ``{Reexamining
  right-handed neutrino EFTs up to dimension six},''
  \href{http://dx.doi.org/10.1103/PhysRevD.106.113008}{{\em Phys. Rev. D}
  {\bfseries 106} no.~11, (2022) 113008},
  \href{http://arxiv.org/abs/2210.12404}{{\ttfamily arXiv:2210.12404
  [hep-ph]}}.

\bibitem{Fernandez-Martinez:2023phj}
E.~Fern\'andez-Mart\'\i{}nez, M.~Gonz\'alez-L\'opez,
  J.~Hern\'andez-Garc\'\i{}a, M.~Hostert, and J.~L\'opez-Pav\'on, ``{Effective
  portals to heavy neutral leptons},''
  \href{http://dx.doi.org/10.1007/JHEP09(2023)001}{{\em JHEP} {\bfseries 09}
  (2023) 001}, \href{http://arxiv.org/abs/2304.06772}{{\ttfamily
  arXiv:2304.06772 [hep-ph]}}.

\bibitem{Beltran:2023ymm}
R.~Beltr\'an, R.~Cepedello, and M.~Hirsch, ``{Tree-level UV completions for
  $N_R$SMEFT $d=6$ and $d=7$ operators},''
  \href{http://dx.doi.org/10.1007/JHEP08(2023)166}{{\em JHEP} {\bfseries 08}
  (2023) 166}, \href{http://arxiv.org/abs/2306.12578}{{\ttfamily
  arXiv:2306.12578 [hep-ph]}}.

\bibitem{Zapata:2023wsz}
G.~Zapata, T.~Urruzola, O.~A. Sampayo, and L.~Duarte, ``{Sensitivity prospects
  for lepton-trijet signals in the $\nu$SMEFT at the LHeC},''
  \href{http://arxiv.org/abs/2305.16991}{{\ttfamily arXiv:2305.16991
  [hep-ph]}}.

\bibitem{Beltran:2023ksw}
R.~Beltr\'an, J.~G\"unther, M.~Hirsch, A.~Titov, and Z.~S. Wang, ``{Heavy
  neutral leptons from kaons in effective field theory},''
  \href{http://arxiv.org/abs/2309.11546}{{\ttfamily arXiv:2309.11546
  [hep-ph]}}.

\bibitem{Gunther:2023vmz}
J.~Y. G\"unther, J.~de~Vries, H.~K. Dreiner, Z.~S. Wang, and G.~Zhou,
  ``{Long-lived neutral fermions at the DUNE near detector},''
  \href{http://arxiv.org/abs/2310.12392}{{\ttfamily arXiv:2310.12392
  [hep-ph]}}.

\bibitem{Deppisch:2015qwa}
F.~F. Deppisch, P.~S. Bhupal~Dev, and A.~Pilaftsis, ``{Neutrinos and Collider
  Physics},'' \href{http://dx.doi.org/10.1088/1367-2630/17/7/075019}{{\em New
  J. Phys.} {\bfseries 17} no.~7, (2015) 075019},
\href{http://arxiv.org/abs/1502.06541}{{\ttfamily arXiv:1502.06541 [hep-ph]}}.

\bibitem{Antusch:2016ejd}
S.~Antusch, E.~Cazzato, and O.~Fischer, ``{Sterile neutrino searches at future
  $e^-e^+$, $pp$, and $e^-p$ colliders},''
  \href{http://dx.doi.org/10.1142/S0217751X17500786}{{\em Int. J. Mod. Phys. A}
  {\bfseries 32} no.~14, (2017) 1750078},
  \href{http://arxiv.org/abs/1612.02728}{{\ttfamily arXiv:1612.02728
  [hep-ph]}}.

\bibitem{Abdullahi:2022jlv}
A.~M. Abdullahi {\em et~al.}, ``{The Present and Future Status of Heavy Neutral
  Leptons},'' in {\em {2022 Snowmass Summer Study}}.
\newblock 3, 2022.
\newblock \href{http://arxiv.org/abs/2203.08039}{{\ttfamily arXiv:2203.08039
  [hep-ph]}}.

\bibitem{Kamp:2022bpt}
N.~W. Kamp, M.~Hostert, A.~Schneider, S.~Vergani, C.~A. Arg\"uelles, J.~M.
  Conrad, M.~H. Shaevitz, and M.~A. Uchida, ``{Dipole-coupled
  heavy-neutral-lepton explanations of the MiniBooNE excess including
  constraints from MINERvA data},''
  \href{http://dx.doi.org/10.1103/PhysRevD.107.055009}{{\em Phys. Rev. D}
  {\bfseries 107} no.~5, (2023) 055009},
  \href{http://arxiv.org/abs/2206.07100}{{\ttfamily arXiv:2206.07100
  [hep-ph]}}.

\bibitem{Abdullahi:2023ejc}
A.~M. Abdullahi, J.~Hoefken~Zink, M.~Hostert, D.~Massaro, and S.~Pascoli, ``{A
  panorama of new-physics explanations to the MiniBooNE excess},''
  \href{http://arxiv.org/abs/2308.02543}{{\ttfamily arXiv:2308.02543
  [hep-ph]}}.

\bibitem{Magill:2018jla}
G.~Magill, R.~Plestid, M.~Pospelov, and Y.-D. Tsai, ``{Dipole Portal to Heavy
  Neutral Leptons},'' \href{http://dx.doi.org/10.1103/PhysRevD.98.115015}{{\em
  Phys. Rev. D} {\bfseries 98} no.~11, (2018) 115015},
  \href{http://arxiv.org/abs/1803.03262}{{\ttfamily arXiv:1803.03262
  [hep-ph]}}.

\bibitem{Schwetz:2020xra}
T.~Schwetz, A.~Zhou, and J.-Y. Zhu, ``{Constraining active-sterile neutrino
  transition magnetic moments at DUNE near and far detectors},''
  \href{http://dx.doi.org/10.1007/JHEP07(2021)200}{{\em JHEP} {\bfseries 21}
  (2020) 200}, \href{http://arxiv.org/abs/2105.09699}{{\ttfamily
  arXiv:2105.09699 [hep-ph]}}.

\bibitem{Brdar:2020quo}
V.~Brdar, A.~Greljo, J.~Kopp, and T.~Opferkuch, ``{The Neutrino Magnetic Moment
  Portal: Cosmology, Astrophysics, and Direct Detection},''
  \href{http://dx.doi.org/10.1088/1475-7516/2021/01/039}{{\em JCAP} {\bfseries
  01} (2021) 039}, \href{http://arxiv.org/abs/2007.15563}{{\ttfamily
  arXiv:2007.15563 [hep-ph]}}.

\bibitem{Brdar:2023tmi}
V.~Brdar, A.~de~Gouv\^ea, Y.-Y. Li, and P.~A.~N. Machado, ``{Neutrino magnetic
  moment portal and supernovae: New constraints and multimessenger
  opportunities},'' \href{http://dx.doi.org/10.1103/PhysRevD.107.073005}{{\em
  Phys. Rev. D} {\bfseries 107} no.~7, (2023) 073005},
  \href{http://arxiv.org/abs/2302.10965}{{\ttfamily arXiv:2302.10965
  [hep-ph]}}.

\bibitem{Ovchynnikov:2022rqj}
M.~Ovchynnikov, T.~Schwetz, and J.-Y. Zhu, ``{Dipole portal and neutrinophilic
  scalars at DUNE revisited: The importance of the high-energy neutrino
  tail},'' \href{http://dx.doi.org/10.1103/PhysRevD.107.055029}{{\em Phys. Rev.
  D} {\bfseries 107} no.~5, (2023) 055029},
  \href{http://arxiv.org/abs/2210.13141}{{\ttfamily arXiv:2210.13141
  [hep-ph]}}.

\bibitem{Ovchynnikov:2023wgg}
M.~Ovchynnikov and J.-Y. Zhu, ``{Search for the dipole portal of heavy neutral
  leptons at future colliders},''
  \href{http://dx.doi.org/10.1007/JHEP07(2023)039}{{\em JHEP} {\bfseries 07}
  (2023) 039}, \href{http://arxiv.org/abs/2301.08592}{{\ttfamily
  arXiv:2301.08592 [hep-ph]}}.

\bibitem{Zhang:2022spf}
Y.~Zhang, M.~Song, R.~Ding, and L.~Chen, ``{Neutrino dipole portal at electron
  colliders},'' \href{http://dx.doi.org/10.1016/j.physletb.2022.137116}{{\em
  Phys. Lett. B} {\bfseries 829} (2022) 137116},
  \href{http://arxiv.org/abs/2204.07802}{{\ttfamily arXiv:2204.07802
  [hep-ph]}}.

\bibitem{Zhang:2023nxy}
Y.~Zhang and W.~Liu, ``{Probing active-sterile neutrino transition magnetic
  moments at LEP and CEPC},''
  \href{http://dx.doi.org/10.1103/PhysRevD.107.095031}{{\em Phys. Rev. D}
  {\bfseries 107} no.~9, (2023) 095031},
  \href{http://arxiv.org/abs/2301.06050}{{\ttfamily arXiv:2301.06050
  [hep-ph]}}.

\bibitem{Liu:2023nxi}
W.~Liu and Y.~Zhang, ``{Testing neutrino dipole portal by long-lived particle
  detectors at the LHC},''
  \href{http://dx.doi.org/10.1140/epjc/s10052-023-11751-0}{{\em Eur. Phys. J.
  C} {\bfseries 83} no.~7, (2023) 568},
  \href{http://arxiv.org/abs/2302.02081}{{\ttfamily arXiv:2302.02081
  [hep-ph]}}.

\bibitem{Barducci:2023hzo}
D.~Barducci, W.~Liu, A.~Titov, Z.~S. Wang, and Y.~Zhang, ``{Probing the dipole
  portal to heavy neutral leptons via meson decays at the high-luminosity
  LHC},'' \href{http://arxiv.org/abs/2308.16608}{{\ttfamily arXiv:2308.16608
  [hep-ph]}}.

\bibitem{Huang:2022pce}
G.-y. Huang, S.~Jana, M.~Lindner, and W.~Rodejohann, ``{Probing heavy sterile
  neutrinos at neutrino telescopes via the dipole portal},''
  \href{http://dx.doi.org/10.1016/j.physletb.2023.137842}{{\em Phys. Lett. B}
  {\bfseries 840} (2023) 137842},
  \href{http://arxiv.org/abs/2204.10347}{{\ttfamily arXiv:2204.10347
  [hep-ph]}}.

\bibitem{Pilaftsis:1991ug}
A.~Pilaftsis, ``{Radiatively induced neutrino masses and large Higgs neutrino
  couplings in the standard model with Majorana fields},''
  \href{http://dx.doi.org/10.1007/BF01482590}{{\em Z. Phys. C} {\bfseries 55}
  (1992) 275--282}, \href{http://arxiv.org/abs/hep-ph/9901206}{{\ttfamily
  arXiv:hep-ph/9901206}}.

\bibitem{BhupalDev:2012zg}
P.~S. Bhupal~Dev, R.~Franceschini, and R.~N. Mohapatra, ``{Bounds on TeV Seesaw
  Models from LHC Higgs Data},''
  \href{http://dx.doi.org/10.1103/PhysRevD.86.093010}{{\em Phys. Rev. D}
  {\bfseries 86} (2012) 093010},
  \href{http://arxiv.org/abs/1207.2756}{{\ttfamily arXiv:1207.2756 [hep-ph]}}.

\bibitem{Cely:2012bz}
C.~G. Cely, A.~Ibarra, E.~Molinaro, and S.~T. Petcov, ``{Higgs Decays in the
  Low Scale Type I See-Saw Model},''
  \href{http://dx.doi.org/10.1016/j.physletb.2012.11.026}{{\em Phys. Lett. B}
  {\bfseries 718} (2013) 957--964},
  \href{http://arxiv.org/abs/1208.3654}{{\ttfamily arXiv:1208.3654 [hep-ph]}}.

\bibitem{Gago:2015vma}
A.~M. Gago, P.~Hernández, J.~Jones-Pérez, M.~Losada, and A.~Moreno~Briceño,
  ``{Probing the Type I Seesaw Mechanism with Displaced Vertices at the LHC},''
  \href{http://dx.doi.org/10.1140/epjc/s10052-015-3693-1}{{\em Eur. Phys. J.}
  {\bfseries C75} no.~10, (2015) 470},
\href{http://arxiv.org/abs/1505.05880}{{\ttfamily arXiv:1505.05880 [hep-ph]}}.

\bibitem{Das:2017zjc}
A.~Das, P.~S.~B. Dev, and C.~S. Kim, ``{Constraining Sterile Neutrinos from
  Precision Higgs Data},''
  \href{http://dx.doi.org/10.1103/PhysRevD.95.115013}{{\em Phys. Rev. D}
  {\bfseries 95} no.~11, (2017) 115013},
  \href{http://arxiv.org/abs/1704.00880}{{\ttfamily arXiv:1704.00880
  [hep-ph]}}.

\bibitem{Maiezza:2015lza}
A.~Maiezza, M.~Nemev\v{s}ek, and F.~Nesti, ``{Lepton Number Violation in Higgs
  Decay at LHC},'' \href{http://dx.doi.org/10.1103/PhysRevLett.115.081802}{{\em
  Phys. Rev. Lett.} {\bfseries 115} (2015) 081802},
  \href{http://arxiv.org/abs/1503.06834}{{\ttfamily arXiv:1503.06834
  [hep-ph]}}.

\bibitem{Deppisch:2018eth}
F.~F. Deppisch, W.~Liu, and M.~Mitra, ``{Long-lived Heavy Neutrinos from Higgs
  Decays},'' \href{http://dx.doi.org/10.1007/JHEP08(2018)181}{{\em JHEP}
  {\bfseries 08} (2018) 181}, \href{http://arxiv.org/abs/1804.04075}{{\ttfamily
  arXiv:1804.04075 [hep-ph]}}.

\bibitem{Accomando:2016rpc}
E.~Accomando, L.~Delle~Rose, S.~Moretti, E.~Olaiya, and C.~H.
  Shepherd-Themistocleous, ``{Novel SM-like Higgs decay into displaced heavy
  neutrino pairs in U(1)' models},''
  \href{http://dx.doi.org/10.1007/JHEP04(2017)081}{{\em JHEP} {\bfseries 04}
  (2017) 081}, \href{http://arxiv.org/abs/1612.05977}{{\ttfamily
  arXiv:1612.05977 [hep-ph]}}.

\bibitem{Mason:2019okp}
J.~D. Mason, ``{Time-Delayed Electrons from Higgs Decays to Right-Handed
  Neutrinos},'' \href{http://dx.doi.org/10.1007/JHEP07(2019)089}{{\em JHEP}
  {\bfseries 07} (2019) 089}, \href{http://arxiv.org/abs/1905.07772}{{\ttfamily
  arXiv:1905.07772 [hep-ph]}}.

\bibitem{ATLAS:2022vhr}
{\bfseries ATLAS} Collaboration, G.~Aad {\em et~al.}, ``{Search for displaced
  photons produced in exotic decays of the Higgs boson using 13~TeV pp
  collisions with the ATLAS detector},''
  \href{http://dx.doi.org/10.1103/PhysRevD.108.032016}{{\em Phys. Rev. D}
  {\bfseries 108} no.~3, (2023) 032016},
  \href{http://arxiv.org/abs/2209.01029}{{\ttfamily arXiv:2209.01029
  [hep-ex]}}.

\bibitem{ATLAS:2013etx}
{\bfseries ATLAS} Collaboration, G.~Aad {\em et~al.}, ``{Search for nonpointing
  photons in the diphoton and $E^{miss}_T$ final state in $\sqrt{s}$=7 TeV
  proton-proton collisions using the ATLAS detector},''
  \href{http://dx.doi.org/10.1103/PhysRevD.88.012001}{{\em Phys. Rev. D}
  {\bfseries 88} no.~1, (2013) 012001},
  \href{http://arxiv.org/abs/1304.6310}{{\ttfamily arXiv:1304.6310 [hep-ex]}}.

\bibitem{ATLAS:2014kbb}
{\bfseries ATLAS} Collaboration, G.~Aad {\em et~al.}, ``{Search for nonpointing
  and delayed photons in the diphoton and missing transverse momentum final
  state in 8 TeV $pp$ collisions at the LHC using the ATLAS detector},''
  \href{http://dx.doi.org/10.1103/PhysRevD.90.112005}{{\em Phys. Rev. D}
  {\bfseries 90} no.~11, (2014) 112005},
  \href{http://arxiv.org/abs/1409.5542}{{\ttfamily arXiv:1409.5542 [hep-ex]}}.

\bibitem{CMS:2012bbi}
{\bfseries CMS} Collaboration, S.~Chatrchyan {\em et~al.}, ``{Search for
  Long-Lived Particles Decaying to Photons and Missing Energy in Proton-Proton
  Collisions at $\sqrt{s}=7$ TeV},''
  \href{http://dx.doi.org/10.1016/j.physletb.2013.04.027}{{\em Phys. Lett. B}
  {\bfseries 722} (2013) 273--294},
  \href{http://arxiv.org/abs/1212.1838}{{\ttfamily arXiv:1212.1838 [hep-ex]}}.

\bibitem{CMS:2019zxa}
{\bfseries CMS} Collaboration, A.~M. Sirunyan {\em et~al.}, ``{Search for
  long-lived particles using delayed photons in proton-proton collisions at
  $\sqrt{s}=$ 13 TeV},''
  \href{http://dx.doi.org/10.1103/PhysRevD.100.112003}{{\em Phys. Rev. D}
  {\bfseries 100} no.~11, (2019) 112003},
  \href{http://arxiv.org/abs/1909.06166}{{\ttfamily arXiv:1909.06166
  [hep-ex]}}.

\bibitem{ATLAS:2023meo}
{\bfseries ATLAS} Collaboration, G.~Aad {\em et~al.}, ``{Search in diphoton and
  dielectron final states for displaced production of Higgs or Z bosons with
  the ATLAS detector in s=13\,\,TeV pp collisions},''
  \href{http://dx.doi.org/10.1103/PhysRevD.108.012012}{{\em Phys. Rev. D}
  {\bfseries 108} no.~1, (2023) 012012},
  \href{http://arxiv.org/abs/2304.12885}{{\ttfamily arXiv:2304.12885
  [hep-ex]}}.

\bibitem{Blennow:2023mqx}
M.~Blennow, E.~Fern\'andez-Mart\'\i{}nez, J.~Hern\'andez-Garc\'\i{}a,
  J.~L\'opez-Pav\'on, X.~Marcano, and D.~Naredo-Tuero, ``{Bounds on lepton
  non-unitarity and heavy neutrino mixing},''
  \href{http://dx.doi.org/10.1007/JHEP08(2023)030}{{\em JHEP} {\bfseries 08}
  (2023) 030}, \href{http://arxiv.org/abs/2306.01040}{{\ttfamily
  arXiv:2306.01040 [hep-ph]}}.

\bibitem{Weinberg:1979sa}
S.~Weinberg, ``{Baryon and Lepton Nonconserving Processes},''
  \href{http://dx.doi.org/10.1103/PhysRevLett.43.1566}{{\em Phys. Rev. Lett.}
  {\bfseries 43} (1979) 1566--1570}.

\bibitem{Anisimov:2006hv}
A.~Anisimov, \href{http://dx.doi.org/10.1142/9789812770288_0058}{``{Majorana
  Dark Matter},''} in {\em 6th International Workshop on the Identification of
  Dark Matter}, pp.~439--449.
\newblock 12, 2006.
\newblock \href{http://arxiv.org/abs/hep-ph/0612024}{{\ttfamily
  arXiv:hep-ph/0612024}}.

\bibitem{Graesser:2007yj}
M.~L. Graesser, ``{Broadening the Higgs boson with right-handed neutrinos and a
  higher dimension operator at the electroweak scale},''
  \href{http://dx.doi.org/10.1103/PhysRevD.76.075006}{{\em Phys. Rev. D}
  {\bfseries 76} (2007) 075006},
  \href{http://arxiv.org/abs/0704.0438}{{\ttfamily arXiv:0704.0438 [hep-ph]}}.

\bibitem{Graesser:2007pc}
M.~L. Graesser, ``{Experimental Constraints on Higgs Boson Decays to TeV-scale
  Right-Handed Neutrinos},'' \href{http://arxiv.org/abs/0705.2190}{{\ttfamily
  arXiv:0705.2190 [hep-ph]}}.

\bibitem{Atre:2009rg}
A.~Atre, T.~Han, S.~Pascoli, and B.~Zhang, ``{The Search for Heavy Majorana
  Neutrinos},'' \href{http://dx.doi.org/10.1088/1126-6708/2009/05/030}{{\em
  JHEP} {\bfseries 0905} (2009) 030},
\href{http://arxiv.org/abs/0901.3589}{{\ttfamily arXiv:0901.3589 [hep-ph]}}.

\bibitem{Kovalenko:2009td}
S.~Kovalenko, Z.~Lu, and I.~Schmidt, ``{Lepton Number Violating Processes
  Mediated by Majorana Neutrinos at Hadron Colliders},''
  \href{http://dx.doi.org/10.1103/PhysRevD.80.073014}{{\em Phys. Rev. D}
  {\bfseries 80} (2009) 073014},
  \href{http://arxiv.org/abs/0907.2533}{{\ttfamily arXiv:0907.2533 [hep-ph]}}.

\bibitem{Helo:2010cw}
J.~C. Helo, S.~Kovalenko, and I.~Schmidt, ``{Sterile neutrinos in lepton number
  and lepton flavor violating decays},''
  \href{http://dx.doi.org/10.1016/j.nuclphysb.2011.07.020}{{\em Nucl. Phys. B}
  {\bfseries 853} (2011) 80--104},
  \href{http://arxiv.org/abs/1005.1607}{{\ttfamily arXiv:1005.1607 [hep-ph]}}.

\bibitem{Bondarenko:2018ptm}
K.~Bondarenko, A.~Boyarsky, D.~Gorbunov, and O.~Ruchayskiy, ``{Phenomenology of
  GeV-scale Heavy Neutral Leptons},''
  \href{http://dx.doi.org/10.1007/JHEP11(2018)032}{{\em JHEP} {\bfseries 11}
  (2018) 032}, \href{http://arxiv.org/abs/1805.08567}{{\ttfamily
  arXiv:1805.08567 [hep-ph]}}.

\bibitem{Coloma:2020lgy}
P.~Coloma, E.~Fern\'andez-Mart\'\i{}nez, M.~Gonz\'alez-L\'opez,
  J.~Hern\'andez-Garc\'\i{}a, and Z.~Pavlovic, ``{GeV-scale neutrinos:
  interactions with mesons and DUNE sensitivity},''
  \href{http://dx.doi.org/10.1140/epjc/s10052-021-08861-y}{{\em Eur. Phys. J.
  C} {\bfseries 81} no.~1, (2021) 78},
  \href{http://arxiv.org/abs/2007.03701}{{\ttfamily arXiv:2007.03701
  [hep-ph]}}.

\bibitem{Branco:1988ex}
G.~C. Branco, W.~Grimus, and L.~Lavoura, ``{The Seesaw Mechanism in the
  Presence of a Conserved Lepton Number},''
  \href{http://dx.doi.org/10.1016/0550-3213(89)90304-0}{{\em Nucl. Phys. B}
  {\bfseries 312} (1989) 492--508}.

\bibitem{Shaposhnikov:2006nn}
M.~Shaposhnikov, ``{A Possible symmetry of the nuMSM},''
  \href{http://dx.doi.org/10.1016/j.nuclphysb.2006.11.003}{{\em Nucl. Phys. B}
  {\bfseries 763} (2007) 49--59},
  \href{http://arxiv.org/abs/hep-ph/0605047}{{\ttfamily arXiv:hep-ph/0605047}}.

\bibitem{Kersten:2007vk}
J.~Kersten and A.~Y. Smirnov, ``{Right-Handed Neutrinos at CERN LHC and the
  Mechanism of Neutrino Mass Generation},''
  \href{http://dx.doi.org/10.1103/PhysRevD.76.073005}{{\em Phys. Rev. D}
  {\bfseries 76} (2007) 073005},
  \href{http://arxiv.org/abs/0705.3221}{{\ttfamily arXiv:0705.3221 [hep-ph]}}.

\bibitem{Gavela:2009cd}
M.~B. Gavela, T.~Hambye, D.~Hernandez, and P.~Hernandez, ``{Minimal Flavour
  Seesaw Models},'' \href{http://dx.doi.org/10.1088/1126-6708/2009/09/038}{{\em
  JHEP} {\bfseries 09} (2009) 038},
  \href{http://arxiv.org/abs/0906.1461}{{\ttfamily arXiv:0906.1461 [hep-ph]}}.

\bibitem{Hernandez:2018cgc}
P.~Hern\'andez, J.~Jones-P\'erez, and O.~Suarez-Navarro, ``{Majorana vs
  Pseudo-Dirac Neutrinos at the ILC},''
  \href{http://dx.doi.org/10.1140/epjc/s10052-019-6728-1}{{\em Eur. Phys. J. C}
  {\bfseries 79} no.~3, (2019) 220},
  \href{http://arxiv.org/abs/1810.07210}{{\ttfamily arXiv:1810.07210
  [hep-ph]}}.

\bibitem{Wyler:1982dd}
D.~Wyler and L.~Wolfenstein, ``{Massless Neutrinos in Left-Right Symmetric
  Models},'' \href{http://dx.doi.org/10.1016/0550-3213(83)90482-0}{{\em Nucl.
  Phys. B} {\bfseries 218} (1983) 205--214}.

\bibitem{Mohapatra:1986bd}
R.~N. Mohapatra and J.~W.~F. Valle, ``{Neutrino Mass and Baryon Number
  Nonconservation in Superstring Models},''
\href{http://dx.doi.org/10.1103/PhysRevD.34.1642}{{\em Phys. Rev.} {\bfseries
  D34} (1986) 1642}.

\bibitem{Malinsky:2005bi}
M.~Malinsky, J.~C. Romao, and J.~W.~F. Valle, ``{Novel supersymmetric SO(10)
  seesaw mechanism},''
  \href{http://dx.doi.org/10.1103/PhysRevLett.95.161801}{{\em Phys. Rev. Lett.}
  {\bfseries 95} (2005) 161801},
  \href{http://arxiv.org/abs/hep-ph/0506296}{{\ttfamily arXiv:hep-ph/0506296}}.

\bibitem{Kang:2006sn}
S.~K. Kang and C.~S. Kim, ``{Extended double seesaw model for neutrino mass
  spectrum and low scale leptogenesis},''
  \href{http://dx.doi.org/10.1016/j.physletb.2006.12.071}{{\em Phys. Lett. B}
  {\bfseries 646} (2007) 248--252},
  \href{http://arxiv.org/abs/hep-ph/0607072}{{\ttfamily arXiv:hep-ph/0607072}}.

\bibitem{Antusch:2017ebe}
S.~Antusch, E.~Cazzato, and O.~Fischer, ``{Resolvable heavy
  neutrino\textendash{}antineutrino oscillations at colliders},''
  \href{http://dx.doi.org/10.1142/S0217732319500615}{{\em Mod. Phys. Lett. A}
  {\bfseries 34} no.~07n08, (2019) 1950061},
  \href{http://arxiv.org/abs/1709.03797}{{\ttfamily arXiv:1709.03797
  [hep-ph]}}.

\bibitem{Fernandez-Martinez:2022gsu}
E.~Fern\'andez-Mart\'\i{}nez, X.~Marcano, and D.~Naredo-Tuero, ``{HNL mass
  degeneracy: implications for low-scale seesaws, LNV at colliders and
  leptogenesis},'' \href{http://dx.doi.org/10.1007/JHEP03(2023)057}{{\em JHEP}
  {\bfseries 03} (2023) 057}, \href{http://arxiv.org/abs/2209.04461}{{\ttfamily
  arXiv:2209.04461 [hep-ph]}}.

\bibitem{Alwall:2014hca}
J.~Alwall, R.~Frederix, S.~Frixione, V.~Hirschi, F.~Maltoni, O.~Mattelaer,
  H.~S. Shao, T.~Stelzer, P.~Torrielli, and M.~Zaro, ``{The automated
  computation of tree-level and next-to-leading order differential cross
  sections, and their matching to parton shower simulations},''
  \href{http://dx.doi.org/10.1007/JHEP07(2014)079}{{\em JHEP} {\bfseries 07}
  (2014) 079},
\href{http://arxiv.org/abs/1405.0301}{{\ttfamily arXiv:1405.0301 [hep-ph]}}.

\bibitem{Buckley:2014ana}
A.~Buckley, J.~Ferrando, S.~Lloyd, K.~Nordstr\"om, B.~Page, M.~R\"ufenacht,
  M.~Sch\"onherr, and G.~Watt, ``{LHAPDF6: parton density access in the LHC
  precision era},''
  \href{http://dx.doi.org/10.1140/epjc/s10052-015-3318-8}{{\em Eur. Phys. J. C}
  {\bfseries 75} (2015) 132}, \href{http://arxiv.org/abs/1412.7420}{{\ttfamily
  arXiv:1412.7420 [hep-ph]}}.

\bibitem{Christensen:2008py}
N.~D. Christensen and C.~Duhr, ``{FeynRules - Feynman rules made easy},''
  \href{http://dx.doi.org/10.1016/j.cpc.2009.02.018}{{\em Comput. Phys.
  Commun.} {\bfseries 180} (2009) 1614--1641},
\href{http://arxiv.org/abs/0806.4194}{{\ttfamily arXiv:0806.4194 [hep-ph]}}.

\bibitem{Alloul:2013bka}
A.~Alloul, N.~D. Christensen, C.~Degrande, C.~Duhr, and B.~Fuks, ``{FeynRules
  2.0 - A complete toolbox for tree-level phenomenology},''
  \href{http://dx.doi.org/10.1016/j.cpc.2014.04.012}{{\em Comput. Phys.
  Commun.} {\bfseries 185} (2014) 2250--2300},
\href{http://arxiv.org/abs/1310.1921}{{\ttfamily arXiv:1310.1921 [hep-ph]}}.

\bibitem{Sjostrand:2006za}
T.~Sjostrand, S.~Mrenna, and P.~Z. Skands, ``{PYTHIA 6.4 Physics and Manual},''
  \href{http://dx.doi.org/10.1088/1126-6708/2006/05/026}{{\em JHEP} {\bfseries
  05} (2006) 026},
\href{http://arxiv.org/abs/hep-ph/0603175}{{\ttfamily arXiv:hep-ph/0603175
  [hep-ph]}}.

\bibitem{Dobbs:2001ck}
M.~Dobbs and J.~B. Hansen, ``{The HepMC C++ Monte Carlo event record for High
  Energy Physics},''
  \href{http://dx.doi.org/10.1016/S0010-4655(00)00189-2}{{\em Comput. Phys.
  Commun.} {\bfseries 134} (2001) 41--46}.

\bibitem{Buckley:2019xhk}
A.~Buckley, P.~Ilten, D.~Konstantinov, L.~L\"onnblad, J.~Monk, W.~Pokorski,
  T.~Przedzinski, and A.~Verbytskyi, ``{The HepMC3 event record library for
  Monte Carlo event generators},''
  \href{http://dx.doi.org/10.1016/j.cpc.2020.107310}{{\em Comput. Phys.
  Commun.} {\bfseries 260} (2021) 107310},
  \href{http://arxiv.org/abs/1912.08005}{{\ttfamily arXiv:1912.08005
  [hep-ph]}}.

\bibitem{LHCHiggsCrossSectionWorkingGroup:2016ypw}
{\bfseries LHC Higgs Cross Section Working Group} Collaboration, D.~de~Florian
  {\em et~al.}, ``{Handbook of LHC Higgs Cross Sections: 4. Deciphering the
  Nature of the Higgs Sector},''
  \href{http://arxiv.org/abs/1610.07922}{{\ttfamily arXiv:1610.07922
  [hep-ph]}}.

\bibitem{FrancoThesis}
F.~Delgado, ``Búsqueda de neutrinos pesados via fotones fuera de tiempo en
  colisionadores,'' bachelor thesis, Pont. U. Catolica, Lima, Peru, February,
  2023.
\newblock Available at \url{http://hdl.handle.net/20.500.12404/24286}.

\bibitem{Nikiforou:2014cka}
N.~Nikiforou, \href{http://dx.doi.org/10.7916/D8668B78}{ ``{Search for
  Non-Pointing Photons in the Diphoton and Missing Transverse Energy Final
  State in 7 TeV $pp$ Collisions Using the ATLAS Detector}''}.
\newblock PhD thesis, Columbia U., 2, 2014.

\bibitem{ATLAS:2009zsq}
{\bfseries ATLAS} Collaboration, G.~Aad {\em et~al.}, ``{Expected Performance
  of the ATLAS Experiment - Detector, Trigger and Physics},''
  \href{http://arxiv.org/abs/0901.0512}{{\ttfamily arXiv:0901.0512 [hep-ex]}}.

\bibitem{deFavereau:2013fsa}
{\bfseries DELPHES 3} Collaboration, J.~de~Favereau, C.~Delaere, P.~Demin,
  A.~Giammanco, V.~Lemaître, A.~Mertens, and M.~Selvaggi, ``{DELPHES 3, A
  modular framework for fast simulation of a generic collider experiment},''
  \href{http://dx.doi.org/10.1007/JHEP02(2014)057}{{\em JHEP} {\bfseries 02}
  (2014) 057},
\href{http://arxiv.org/abs/1307.6346}{{\ttfamily arXiv:1307.6346 [hep-ex]}}.

\bibitem{Cacciari:2011ma}
M.~Cacciari, G.~P. Salam, and G.~Soyez, ``{FastJet User Manual},''
  \href{http://dx.doi.org/10.1140/epjc/s10052-012-1896-2}{{\em Eur. Phys. J.}
  {\bfseries C72} (2012) 1896},
\href{http://arxiv.org/abs/1111.6097}{{\ttfamily arXiv:1111.6097 [hep-ph]}}.

\bibitem{ATLAS:2008xda}
{\bfseries ATLAS} Collaboration, G.~Aad {\em et~al.}, ``{The ATLAS Experiment
  at the CERN Large Hadron Collider},''
  \href{http://dx.doi.org/10.1088/1748-0221/3/08/S08003}{{\em JINST} {\bfseries
  3} (2008) S08003}.

\bibitem{ATLAS:2019qmc}
{\bfseries ATLAS} Collaboration, G.~Aad {\em et~al.}, ``{Electron and photon
  performance measurements with the ATLAS detector using the
  2015\textendash{}2017 LHC proton-proton collision data},''
  \href{http://dx.doi.org/10.1088/1748-0221/14/12/P12006}{{\em JINST}
  {\bfseries 14} no.~12, (2019) P12006},
  \href{http://arxiv.org/abs/1908.00005}{{\ttfamily arXiv:1908.00005
  [hep-ex]}}.

\bibitem{ATLAS:2020auj}
{\bfseries ATLAS} Collaboration, G.~Aad {\em et~al.}, ``{Muon reconstruction
  and identification efficiency in ATLAS using the full Run 2 $pp$ collision
  data set at $\sqrt{s}=13$ TeV},''
  \href{http://dx.doi.org/10.1140/epjc/s10052-021-09233-2}{{\em Eur. Phys. J.
  C} {\bfseries 81} no.~7, (2021) 578},
  \href{http://arxiv.org/abs/2012.00578}{{\ttfamily arXiv:2012.00578
  [hep-ex]}}.

\bibitem{Read:2002hq}
A.~L. Read, ``{Presentation of search results: The CL(s) technique},''
  \href{http://dx.doi.org/10.1088/0954-3899/28/10/313}{{\em J. Phys.}
  {\bfseries G28} (2002) 2693--2704}.
[,11(2002)].

\bibitem{ParticleDataGroup:2022pth}
{\bfseries Particle Data Group} Collaboration, R.~L. Workman {\em et~al.},
  ``{Review of Particle Physics},''
  \href{http://dx.doi.org/10.1093/ptep/ptac097}{{\em PTEP} {\bfseries 2022}
  (2022) 083C01}.

\bibitem{CMS:2018uag}
{\bfseries CMS} Collaboration, A.~M. Sirunyan {\em et~al.}, ``{Combined
  measurements of Higgs boson couplings in proton\textendash{}proton collisions
  at $\sqrt{s}=13\,\text {Te}\text {V} $},''
  \href{http://dx.doi.org/10.1140/epjc/s10052-019-6909-y}{{\em Eur. Phys. J. C}
  {\bfseries 79} no.~5, (2019) 421},
  \href{http://arxiv.org/abs/1809.10733}{{\ttfamily arXiv:1809.10733
  [hep-ex]}}.

\bibitem{ATLAS:2019nkf}
{\bfseries ATLAS} Collaboration, G.~Aad {\em et~al.}, ``{Combined measurements
  of Higgs boson production and decay using up to $80$ fb$^{-1}$ of
  proton-proton collision data at $\sqrt{s}=$ 13 TeV collected with the ATLAS
  experiment},'' \href{http://dx.doi.org/10.1103/PhysRevD.101.012002}{{\em
  Phys. Rev. D} {\bfseries 101} no.~1, (2020) 012002},
  \href{http://arxiv.org/abs/1909.02845}{{\ttfamily arXiv:1909.02845
  [hep-ex]}}.

\bibitem{Aparici:2009oua}
A.~Aparici, A.~Santamaria, and J.~Wudka, ``{A model for right-handed neutrino
  magnetic moments},''
  \href{http://dx.doi.org/10.1088/0954-3899/37/7/075012}{{\em J. Phys. G}
  {\bfseries 37} (2010) 075012},
  \href{http://arxiv.org/abs/0911.4103}{{\ttfamily arXiv:0911.4103 [hep-ph]}}.

\end{thebibliography}\endgroup

\end{document}